\documentclass[
reprint,
longbibliography,
superscriptaddress,
amsmath,amssymb,
aps,
prb,
]{revtex4-2}

\usepackage{amsfonts}

\usepackage{graphicx}
\usepackage{dcolumn}
\usepackage{bm}
\usepackage{hyperref}
\usepackage{physics}
\usepackage{xcolor}
\usepackage{mathrsfs}
\usepackage{mathtools}
\usepackage{comment}
 \usepackage{lipsum}
 \usepackage{amsfonts}
 \usepackage{bbm}
 \usepackage{leftindex}

\usepackage[american,british]{babel}
\usepackage[T1]{fontenc}
\usepackage{dcolumn}
\usepackage{braket}
\usepackage{amsmath,amsthm,amssymb}
\usepackage{color}
\usepackage{verbatim}
\usepackage{lipsum}

\usepackage{dsfont}
\usepackage{physics}
\usepackage{soul}

\begin{document}

\title{Dynamical deconfinement transition driven by density of excitations}

\author{Nishan Ranabhat}
\email{nranabha@sissa.it}
\affiliation{SISSA, via Bonomea 265, 34136 Trieste, Italy}
\affiliation{ICTP, Strada Costiera 11, 34151 Trieste, Italy}

\author{Alessandro Santini}
\email{asantini@sissa.it}
\affiliation{SISSA, via Bonomea 265, 34136 Trieste, Italy}

\author{Emanuele Tirrito}
\email{emanuele.tirrito@unitn.it}
\affiliation{ICTP, Strada Costiera 11, 34151 Trieste, Italy}
\affiliation{Pitaevskii BEC Center, CNR-INO and Dipartimento di Fisica, Università di Trento, Via Sommarive 14, Trento, I-38123, Italy}

\author{Mario Collura}
\email{mcollura@sissa.it}
\affiliation{SISSA, via Bonomea 265, 34136 Trieste, Italy}
\affiliation{INFN, via Bonomea 265, 34136 Trieste, Italy}

\date{\today}

\begin{abstract}
We investigate the dynamical deconfinement transition driven by excitations in long-range Ising model. At low temperatures, spatially separated pairs of domain wall kinks are bound by the confining potential and exhibit uncorrelated Bloch oscillations in time. This picture is analogous to bound mesons in quark confinement. As the temperature increases, the meson picture breaks down as the domain wall kinks in proximity interact and disperse, leading to an extended deconfined regime. In this study, we characterize the deconfinement transition with signatures observed in the average density of domain wall kinks and nonequilibrium changes in its fluctuation. Our findings provide insights into the mechanisms of confinement and deconfinement in long-range spin models, thus opening avenues for further exploration and experimental verification.
\end{abstract}

\maketitle

\noindent \textbf{Introduction}:
Confinement, as observed in quantum chromodynamics (QCD), is a phenomenon that binds fundamental particles, such as quarks, into stable heavier particles known as hadrons \cite{coleman2015introduction,greiner2007quantum,busza2018heavy,berges2021qcd,rothkopf2020heavy}. This binding occurs because of the presence of a confining potential, which increases asymptotically with particle separation. Recent research in lattice gauge theories \cite{LGT_CONF1,LGT_CONF2,LGT_CONF3,LGT_CONF4,LGT_CONF5,LGT_CONF6,LGT_CONF7,zhang2023observation,PRXQuantum.3.040316} and quantum spin chains \cite{LRIM_CONF_1,LRIM_CONF_2,LRIM_CONF_expt,SR_conf,SR_CONF_expt,PhysRevB.104.L201106,lagnese2022quenches} has explored the effects of confinement in the out-of-equilibrium dynamics, revealing anomalous dynamical behavior signified by suppressed correlation and entanglement growth, and slow thermalization \cite{LGT_CONF1,LRIM_CONF_2,LRIM_CONF_1,LRIM_CONF_expt,ourown,ranabhat2023thermalization,SR_conf,SR_CONF_expt,Alvise_conf_entang,PRETHERMAL_1}. Despite the fundamental differences between the lattice gauge theories and quantum spin chains, there is a connection between the observed confinement behavior in these two systems. Notably, the two-dimensional Ising model can exhibit confinement-like behavior, similar to that observed in lattice gauge theories \cite{belavin1984infinite,henkel1989two,fonseca2006ising}. Therefore, the study of simplified magnetic spin models can provide valuable insights into the mechanisms of confinement and phase transitions in complex gauge theories \cite{meson_melting}.

In recent years, there has been increasing interest in the study of confinement-deconfinement transitions in lattice gauge theories~\cite{Jadconfinement,LGT_CONF7,PhysRevX.6.041040,aidelsburger2022cold,buazuavan2023synthetic,mildenberger2022probing,LGT_deconf_1,LGT_deconf_2} and quantum spin chains~\cite{SR_deconf_Alessio,LRIM_CONF_expt,mcbanuls}. This interest is driven by the direct analogy between this phenomenon and the confinement-deconfinement crossover proposed in QCD, where stable hadrons transition into quark-gluon plasma at exceedingly high energy densities. Previous studies have demonstrated a confinement-deconfinement transition in the short-range Ising model with transverse and longitudinal field, characterized by logarithmic entanglement entropy growth in the confined regime, transitioning to linear growth in the deconfined regime \cite{SR_deconf_Alessio}. Furthermore, it has been observed that in high energy density regimes, rare events with mesons created in close proximity lead to avalanches of scattering events, resulting in a stable prethermal regime even in the presence of confinement \cite{birnkammer2022prethermalization}. Moreover, in the long-range Ising model (LRIM)~\cite{Defenu2023Review_1,Defenu2023Review_2}, a stable deconfined regime has been observed in a system of trapped ions at a high transverse field \cite{LRIM_CONF_expt}.

\begin{figure*}[t!]
    \centering
    \includegraphics[width=0.8\linewidth]{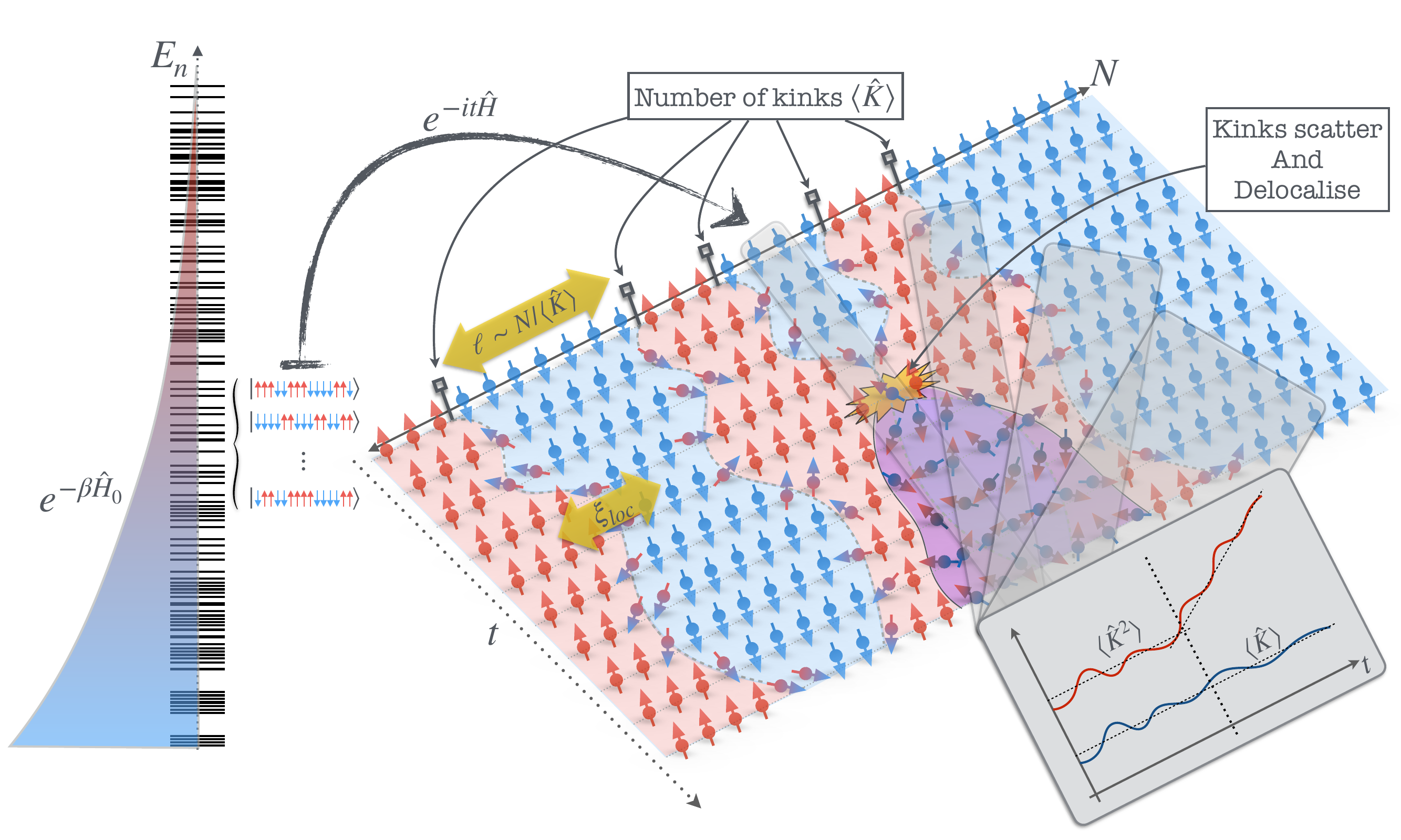}
    \caption{Schematic diagram of the quench protocol: \textbf{1)} Initial states are prepared as thermal density matrices of the initial Hamiltonian, $\propto e^{-\beta \hat{H}_0}$ at varying temperatures ($\beta = 1/T$). \textbf{2)} The initial thermal density matrix is evolved in real time with a final Hamiltonian $\propto e^{-i t \hat{H}}$ . Here, we illustrate the evolution of a single representative state withing a thermal density matrix. $\langle K \rangle$ is the average number of domain wall kinks, $l$ is the average distance between two kinks and $\xi_{loc}$ is the localization length defined as the maximum distance traced by a domain wall kink. \textbf{3)} We probe the real time evolution by calculating the kink density and kink fluctuation at every time step.}
    \label{fig:schematic_quench_protocol}
\end{figure*}

In this study, we investigate the robustness of confinement behavior in LRIM against the thermally tuned density of excitations. The long-range power-law decaying interaction in LRIM naturally induces confinement between a pair of domain-wall kinks \cite{LRIM_CONF_1,LRIM_CONF_2,ranabhat2023thermalization}. We begin with a thermal state with a finite initial kink density that is tuned by the temperature, followed by a quench to the strongly confining regime \cite{LRIM_CONF_2,LRIM_CONF_expt,Alvise_conf_entang} and investigate the real-time dynamics of the average kink density and kink fluctuation \cite{LRIM_CONF_expt,SR_CONF_expt}. We employ state-of-the-art tensor network simulations \cite{schollwock2011density,orus2014practical,ran2020tensor} to represent the thermal states as matrix density operators (MPDO) \cite{fintem1,fintem2,LDCarr} and simulate their subsequent real-time evolution. The post-quench real-time dynamics of a mixed density matrix exhibit a much richer behavior compared to that of a pure state with a non-zero density of kinks. Both average kink density and kink fluctuation show strong signatures of transition from a low-temperature strongly confined phase to a high-temperature deconfined phase. Finally, we employ an effective semi-classical model with a single kink \cite{LRIM_CONF_1} to theoretically predict the observed confinement-deconfinement transition. Our results can be experimentally realized using various Atomic, Molecular, and Optical (AMO) platforms \cite{DYN_PT_expt_1,DYN_PT_expt_2,LRIM_CONF_expt,LRIM_EXPT1,LRIM_EXPT2,LRIM_EXPT3,LRIM_EXPT4,LRIM_EXPT5,LRIM_EXPT6} that are capable of implementing the global quench protocol.\\

\noindent \textbf{Model:}
\noindent We study a spin-1/2 ferromagnetic long-range Ising model (LRIM) under open boundary conditions, described by the following Hamiltonian:

\begin{equation}
\hat{H}(J,\alpha,h) = -J\sum_{i<j}^N \frac{\hat{\sigma}^x_i \hat{\sigma}^x_j }{|i-j|^{\alpha}}- h \sum_{i=1}^N \hat{\sigma}^z_i
\label{eq:LRIM}
\end{equation}

Here, $\hat{\sigma}_i^{\nu}$ represents the Pauli matrix in the $i^{th}$ site and $\nu^{th}$ directions. The long-range interaction between the two spins follows an inverse power law of their separation distance, controlled by the parameter $\alpha$. The transverse magnetic field $h$ governs the kinetic term in the Hamiltonian. We adopt a energy scaling convention with $J$ set to 1. The LRIM demonstrates integrable behavior at the two extremities of $\alpha$. At $\alpha = \infty$, it simplifies to the nearest-neighbor transverse field Ising model (TFIM) and can be effectively solved by mapping onto spinless fermions through the Jordan-Wigner transformation \cite{TFIM_JW}. Conversely, when $\alpha = 0$, it transforms into a fully connected model \cite{FC1,FC2,ourown}. TFIM exhibits a quantum phase transition between the ferromagnetic and paramagnetic phases at $h=J$, which persists for all values of $\alpha$ with an increasing critical point \cite{QPT1,QPT2,QPT3}. Additionally, at $\alpha \leq 2$ we observe a thermal phase transition in the LRIM from a low-temperature ferromagnetic phase to a high-temperature paramagnetic phase  \cite{LRIM_T_trans1,LRIM_T_trans2}. In recent years, LRIM has emerged as a paradigmatic spin model for studying nonequilibrium physics in many-body systems owing to its rich array of intriguing features, including dynamical phase transitions \cite{ourown,DYN_PT_1,DYN_PT_2,DYN_PT_3,DYN_PT_expt_1,DYN_PT_expt_2,Halimeh_2,Halimeh_3}, prethermalization phenomena \cite{Halimeh_1,PRETHERMAL_1}, nonlinear lightcone propagation \cite{LIGHT_CONE_1,LIGHT_CONE_2}, and the emergence of time crystals \cite{DTC1,DTC2}. Furthermore, LRIM exhibits the phenomenon of dynamical confinement, wherein it confines pairs of domain wall kinks into bound quasiparticles, ultimately resulting in the suppression of information propagation within the system \cite{LRIM_CONF_1,LRIM_CONF_2,LRIM_CONF_expt,Alvise_conf_entang}.\\

\noindent \textbf{Quench dynamics}: We initialize our states as a thermal density matrix $\hat{\rho}_{\beta}(t=0) \propto e^{-\beta \hat{H}_0}$, where $\beta$ represents the inverse temperature and the initial Hamiltonian $\hat{H}_0$ has a zero magnetic field. The initial states evolve in real time with the post-quench Hamiltonian  $\hat{\rho}_{\beta}(t+dt) = e^{-i dt \hat{H}}\hat{\rho}_{\beta}(t)e^{i dt \hat{H}}$ (see figure \ref{fig:schematic_quench_protocol}). The post-quench Hamiltonian parameters are selected to reside within the strongly confined regime \cite{LRIM_CONF_1,LRIM_CONF_2}. The thermal state is simulated as a mixed density matrix in a locally purified form \cite{fintem1,fintem2}. We begin at an infinite temperature, that is, $\beta = 0$ where the state is maximally entangled and has a trivial representation as the matrix product density operator (MPDO)\cite{fintem1}. We obtain subsequent finite temperature states by the imaginary time evolution of the maximally mixed state. The thermal states realized as MPDO are then evolved in real time. We employ a two-site time dependent variational algorithm (TDVP) \cite{TDVP_one,TDVP_two} with second order integration scheme for both imaginary and real time evolution of MPDO.  (See Appendix) for further details on the numerical simulation. 

The elementary excitations of the dynamics are domain wall kinks, which propagate through the system in real time spreading correlation. Consequently, kink density emerges as a suitable order parameter for investigating confinement and its impact on correlation propagation within the system \cite{LRIM_CONF_1, LRIM_CONF_2, LRIM_CONF_expt, SR_conf}. Specifically, it is defined as
\begin{equation}
\hat{k} = \frac{1}{N}\sum_{i=1}^{N-1} \frac{1-\hat{\sigma}^x_i
\hat{\sigma}^x_{i+1}}{2},
\label{eq:kink_den_oprt}
\end{equation}
and quantifies the number of nearest neighbor kinks within the system. The initial state is represented by a mixed thermal density matrix, with the average kink density ranging from $0.5$ at infinite temperature (corresponding to a maximally mixed state) to $0$ at lower temperatures (see Appendix). Therefore, the parameter $\beta$ serves as a tuning parameter to precisely control the initial state of the system with varying kink densities.

\begin{figure}[t!]
    \centering
    \includegraphics[width=1.00\linewidth]{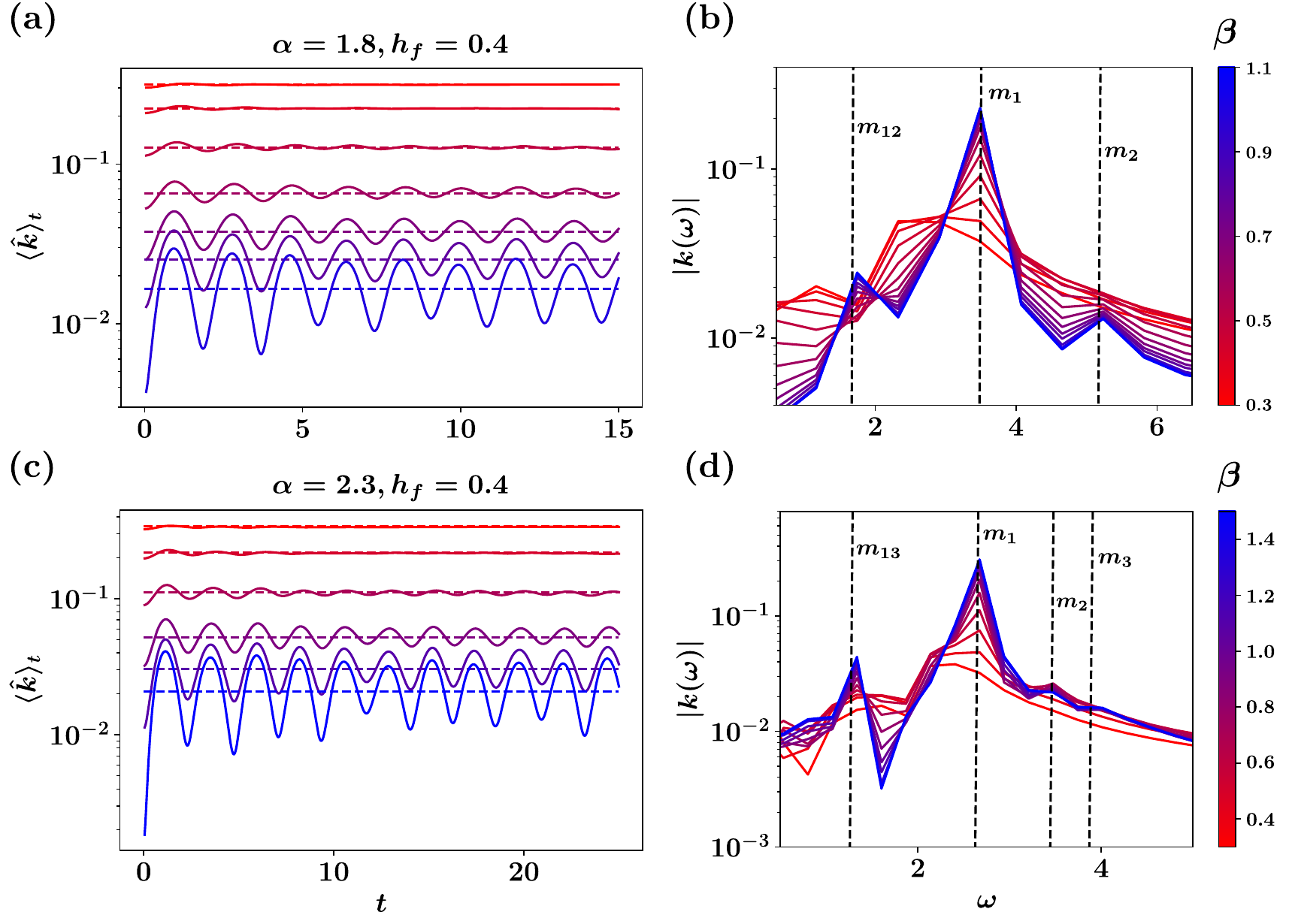}
    \caption{Post-quench evolution of kink density starting from different thermal states (see colorbars on the right) for $\alpha = 1.8, h_f = 0.4$ \textbf{(a)} and $\alpha = 2.3, h_f = 0.4$ \textbf{(c)}. The horizontal dashed lines represent the kink densities of the thermal states corresponding to the quenches. The results are for system size of $N=50$ spins. Panels \textbf{(b)} and \textbf{(c)} are the corresponding Fourier spectrum. The vertical dashed lines corresponds to the meson masses and their differences calculated from the effective two kink model, see appendix} 
    \label{fig:kink_avg_meson_mass}
\end{figure}

In Figure \ref{fig:kink_avg_meson_mass} panels \textbf{(a)} and \textbf{(c)}, we plot the post-quench evolution of the average kink density,  $\langle \hat{k} \rangle$ for two representative post-quench parameters, initiated from various initial thermal states. The horizontal dashed lines represent the expected thermal values $\text{Tr}[\hat{\rho}_{\tilde{\beta}}\hat{k}]$, where $\hat{\rho}_{\tilde{\beta}} \propto e^{-\tilde{\beta}\hat{H}}$ and the effective temperature attributed to a quench \cite{Essler_2016} $\tilde{\beta}$ is extracted by solving the equation,

\begin{equation}
\frac{\text{Tr}[\hat{\rho}_{\beta}\hat{H}_0]}{\text{Tr}[\hat{\rho}_{\beta}]} = \frac{\text{Tr}[\hat{\rho}_{\tilde{\beta}}\hat{H}]}{\text{Tr}[\hat{\rho}_{\tilde{\beta}}]}
\label{eq:eff_temp}
\end{equation}

\begin{figure}[t!]
    \centering
    \includegraphics[width=0.8\linewidth]{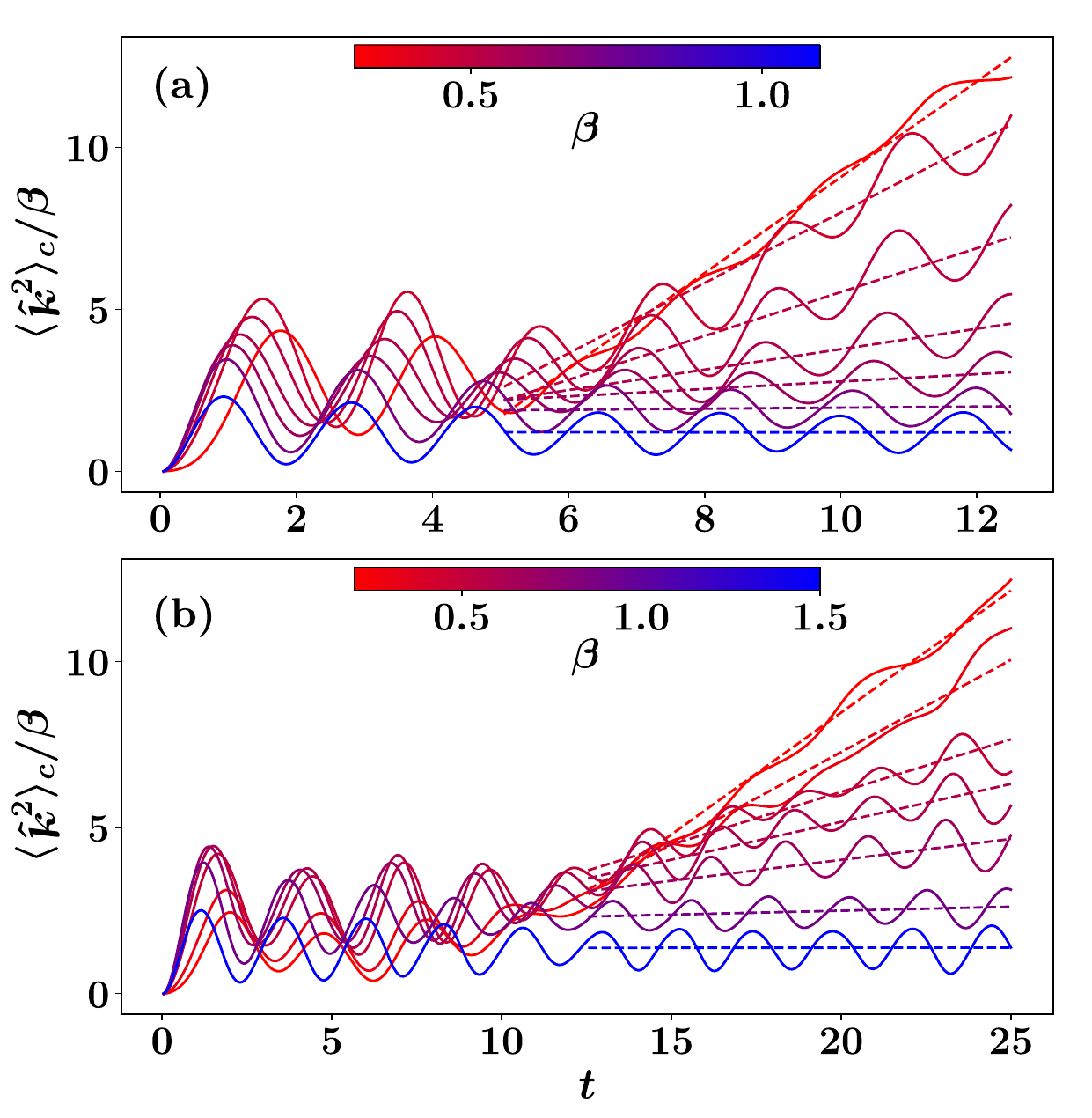}
    \caption {Time evolution of the kink fluctuation at different temperatures (the color bar shows the inverse temperature $\beta$) for $\alpha = 1.8, h_f = 0.4$ \textbf{(a)} and $\alpha = 2.3, h_f = 0.4$ \textbf{(b)}.The results are for system size of $N=50$ spins. The data are rescaled by $1/\beta$ and subtracted from their initial values for better visualization. Dashed lines represent the linear fit of the data in an appropriate time window.} 
    \label{fig:kink_fluctuate_LRIM}
\end{figure}

We observe two distinct behaviors at the extremities of the investigated temperature range. At lower temperatures, the kink density persistently oscillates around the expected thermal value. This is indicative of strong confinement and suppression of the correlation spreading within the system and eventual delay of thermalization \cite{SandW_therm}. This behavior weakens as the temperature increases, and at high temperatures, the kink density rapidly relaxes to the expected thermal value, suggesting robust thermalization \cite{SandW_therm} and deconfinement \cite{SR_conf,LRIM_CONF_2}. Figure \ref{fig:kink_avg_meson_mass} panels \textbf{(b)} and \textbf{(d)} show the corresponding Fourier spectrum of $\langle \hat{k} \rangle$.
At lower temperatures we observe sharp frequency peaks corresponding to the dominant oscillations of $\langle \hat{k} \rangle$. These frequency peaks exhibit strong agreement with the meson masses extracted from the two-kink model \cite{SR_conf,SR_CONF_expt,LRIM_CONF_1,LRIM_CONF_2,Alvise_conf_entang} (see Appendix for additional details) indicative of the presence of strongly bound mesons. However, as the temperature increases, these frequency peaks gradually diminish and ceases to exists, indicating the dissolution of bound mesons at high temperature \cite{meson_melting}.\\

\noindent \textbf{Evolution of kink fluctuation}:
Kink density provides valuable qualitative insights into the deconfinement transition, revealing distinct behaviors at the extreme ends of the temperature range under examination. However, a more prominent signature of this transition emerges from the interaction and propagation of domain wall kinks within the system. To address this, we study the connected kink fluctuation
$\langle \hat{k}^2 \rangle_c = \langle \hat{k}^2 \rangle
- \langle \hat{k} \rangle^2$.

\begin{figure}[t!]
    \centering
    \includegraphics[width=1.00\linewidth]{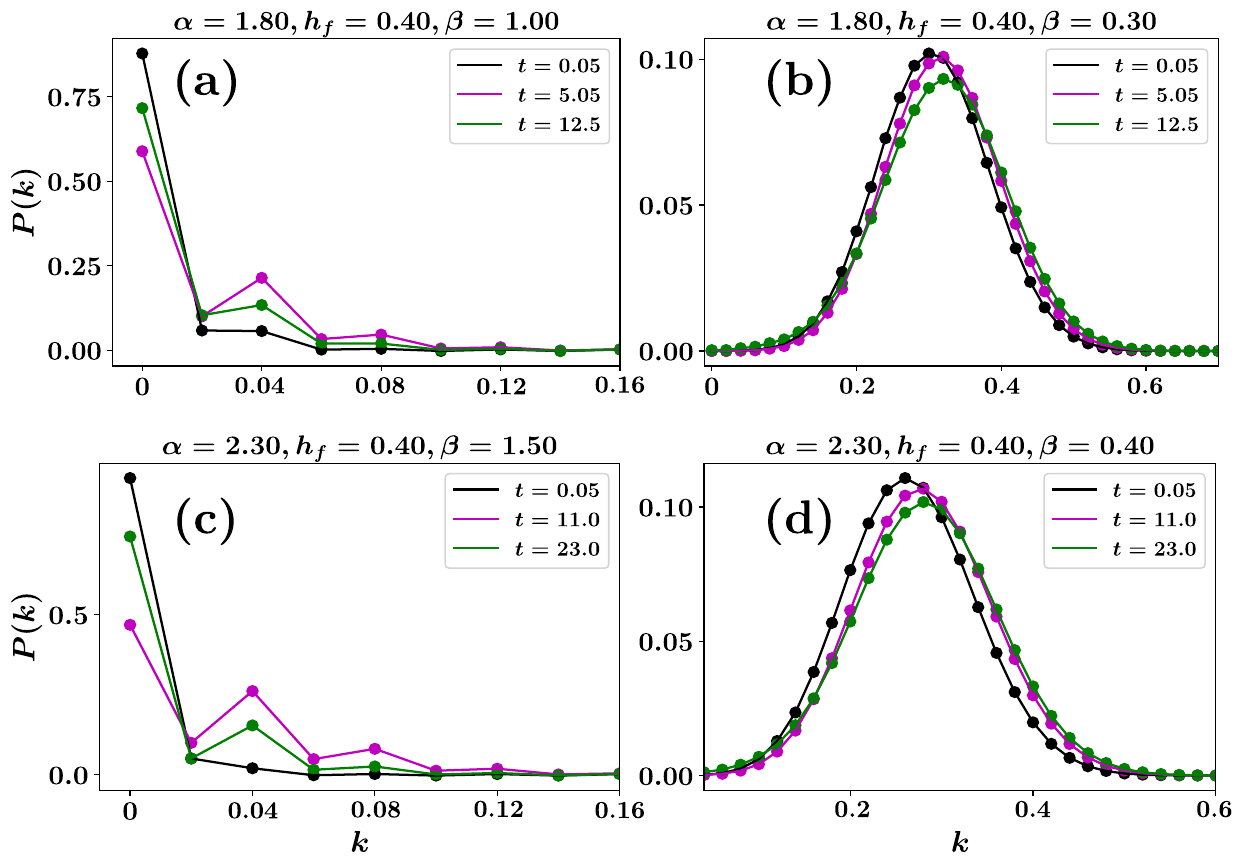}
    \caption{Probability distribution function (PDF) of domain wall kinks for different system parameters and temperature. Three colors represent different time slices during the real time dynamics (see label). \textbf{(a)} and \textbf{(c)} exhibits strong confinement at low temperature with persistent oscillation of PDF, \textbf{(b)} and \textbf{(d)} exhibits high temperature deconfinement where the PDF consistently gets broader with time (see Appendix for details on calculation of $P(k)$)}.
    \label{fig:PDF_kink}
\end{figure}

\begin{figure}[t!]
    \centering
    \includegraphics[width=1.00\linewidth]{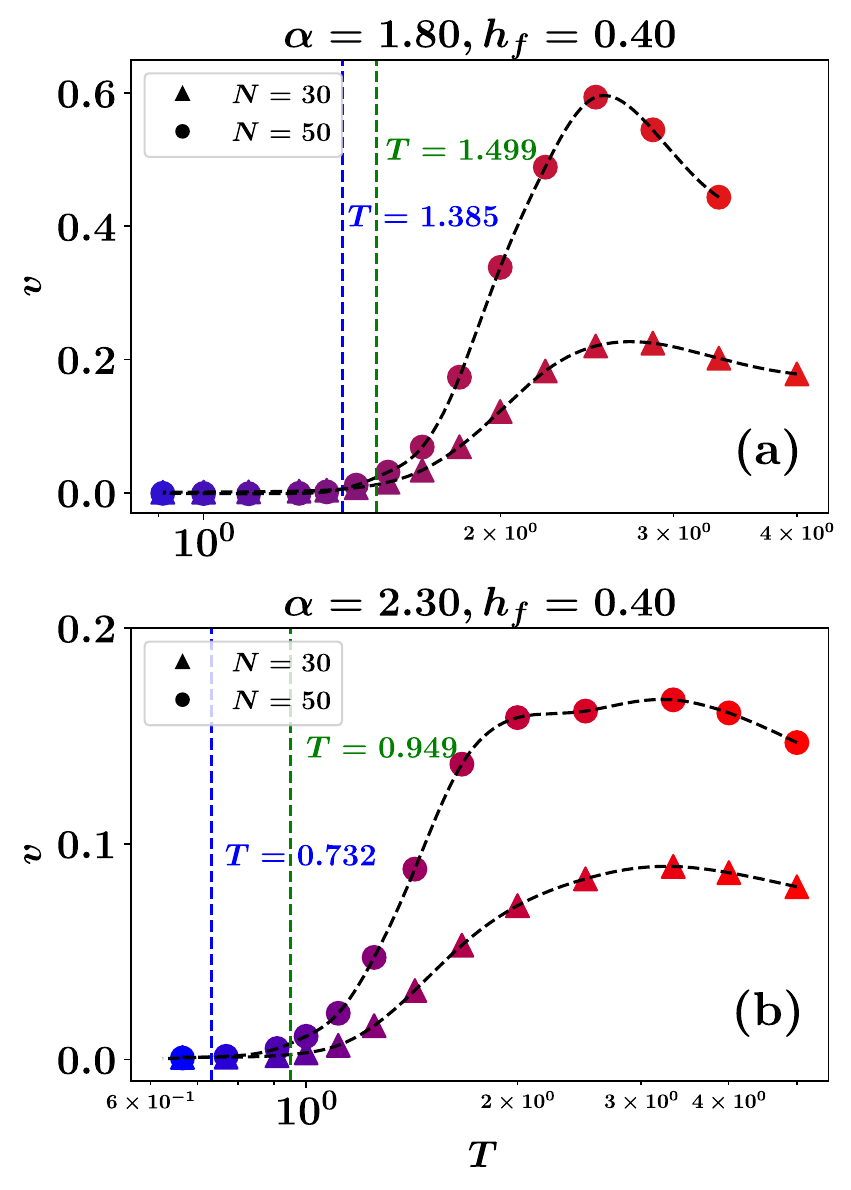}
    \caption{Velocity of kink fluctuation extracted from \ref{fig:kink_fluctuate_LRIM} as a function temperature for two different quenches: $\alpha=1.8,h_f=0.4$ \textbf{(a)} and $\alpha=2.3,h_f=0.4$ \textbf{(b)} and system sizes $N=30$ (solid triangle) and $N=50$ (solid circle). The color gradient represents varying temperature. Dashed black lines are for visual guidance. Horizontal dashed lines are the deconfinement transition temperatures predicted by the single kink model: blue for $N=50$, green for $N=30$.} 
    \label{fig:slope_temp_alpha}
\end{figure}

In Figure \ref{fig:kink_fluctuate_LRIM}, we depict the post-quench evolution of $\langle \hat{k}^2 \rangle_c$ for two distinct sets of post-quench parameters. For both post-quench Hamiltonians, we observe two contrasting behaviors at the extremes of the temperature range considered. At lower temperatures, $\langle \hat{k}^2 \rangle_c$ exhibits persistent oscillations over time, similar to the behavior of $\langle \hat{k} \rangle$, indicating a significant suppression in the propagation of domain wall kinks. As the temperature increases, $\langle \hat{k}^2 \rangle_c$ exhibits linear growth over time, suggesting a light-cone-like dispersion of the domain wall kinks. This observation aligns with light-cone spreading of correlation and entanglement in the deconfined phase, as previously discussed in the literature \cite{LRIM_CONF_1,LRIM_CONF_2,SR_conf}. A more comprehensive understanding emerges when we examine the full probability distribution $P(k)$ of the domain wall kinks. In Figure \ref{fig:PDF_kink} panels \textbf{(b)} and \textbf{(d)}, we observe that $P(k)$ monotonically broadens over time, whereas in panels \textbf{(a)} and \textbf{(c)}, it exhibits oscillations in time. To quantitatively assess this behavior, we fit straight lines to the $\langle \hat{k}^2 \rangle_c$ data over an appropriate time window and define the slope of these lines as the velocity of kink dispersion $v$. In figure \ref{fig:slope_temp_alpha} we plot $v$ as the function of temperature of the initial thermal states. We observe a transition from a strongly confined regime at low temperatures denoted by $v \approx 0$ to a deconfined regime denoted by a monotonically increasing finite $v$ as the temperature increases. This monotonic rise in $v$ halts eventually because of the finite size effects.\\

The underlying mechanism of this transition can be understood by studying the dynamics of an effective single kink model, previously employed to study the quasi-localized excitations and suppression of transport in one-dimensional spin chains \cite{LRIM_CONF_1,single_kink1}. The semi-classical limit of the single kink model is given by the Hamiltonian,

\begin{equation}
    H^{\text{cls}}_{\alpha,h,N}(k,q) = V_{\alpha,N}(q) - 2h \hspace{0.1cm} \text{cos}(k)
\label{eq:1kink_semiclassical_mt}
\end{equation}

where,

\begin{align}
    \begin{split}
    V_{\alpha,N}(q) 
    = \frac{2[q^{2-\alpha} + (N-q)^{2-\alpha}-
    (N-1)^{2-\alpha}]}{(\alpha-1)(2-\alpha)},
    \end{split}
\label{eq:1kink_pot_approx_mt}
\end{align}

and $(k,q) \in [0,2\pi]\times [1,N-1]$ (see Appendix for further details). Within this framework we define the localization length for a given system parameters, $\xi_{loc}$ as the maximum distance traced by the single kink initialized at rest with the maximum possible energy. This is extracted by solving for maximum $q$ in the equation,

\begin{equation}
    H^{\text{cls}}_{\alpha,N}(0,N/2) =  H^{\text{cls}}_{\alpha,N}(k,q).
\label{eq:phase_sol_mt}
\end{equation}

$\xi_{loc}$ provides a natural threshold for deconfinement transition. When the separation between kinks is larger than localization length, $ l = 1/\langle \hat{k} \rangle \gg \xi_{loc}$ the meson picture remains valid and pair of domain wall kinks undergo uncorrelated Bloch oscillations localized at their site of origin. However, when $ 1/\langle \hat{k} \rangle \lessapprox \xi_{loc}$, the meson picture breaks down and kinks interact and scatter, effectively destroying confinement \cite{birnkammer2022prethermalization}. We can extract the corresponding transition temperature by solving the following equation for $\beta$,
\begin{equation}
    \frac{1}{\xi_{loc}} = \overline{\langle \hat{k}_{\beta}(t) \rangle},
\label{eq:extract_T_crit}
\end{equation}
where $\overline{\langle \hat{k}_{\beta}(t) \rangle}$ is the time averaged $\text{Tr}[\hat{\rho}_{\beta}(t)\hat{k}]$. It is apparent from Figure \ref{fig:kink_avg_meson_mass} that $\overline{\langle \hat{k}_{\beta}(t) \rangle} = \text{Tr}[\hat{\rho}_{\tilde{\beta}}\hat{k}]$ holds true for all values of $\beta$. By substituting this relation into Equation \ref{eq:extract_T_crit}, the transition temperature can be determined numerically. In Figure \ref{fig:slope_temp_alpha}, the dashed horizontal lines represent the deconfinement transition temperatures obtained using this method. Despite the simplicity of the single-kink model, it demonstrates a strong predictive capacity for transition temperature.\\
This observation can be realized in trapped-ion experiments and other Atomic, Molecular, and Optical (AMO) platforms \cite{DYN_PT_expt_1,DYN_PT_expt_2,LRIM_CONF_expt,LRIM_EXPT1,LRIM_EXPT2,LRIM_EXPT3,LRIM_EXPT4,LRIM_EXPT5,LRIM_EXPT6} that are capable of initiating the global quench protocol from an initial product state. The post-quench evolution of a mixed state can be realized by independently evolving individual pure states within the given mixed state and subsequently computing the weighted ensemble average of the individual pure state evolution: $\hat{\rho}_{\beta}(0) \rightarrow \hat{\rho}_{\beta}(t) = \sum_{n} P_{\beta}(n) \ket{n(t)} \bra{n(t)}$, where $\ket{n(t)} = e^{-i t \hat{H}} \ket{n}$ and $P_{\beta}(n) = e^{-\beta E_n} / \sum_n e^{-\beta E_n}$. State $\ket{n}$ represents the eigenstate of $\hat{H}_0$, and $E_n$ corresponds to the associated eigenvalue. Notably, this procedure scales exponentially with the system size. However, for practical purposes, it is feasible to consider only the dominant states, based on how $P_{\beta}(n)$ decays  with $n$. This is particularly applicable to low-temperature states.\\

\noindent \textbf{Conclusion}: 
We have studied the out-of -equilibrium dynamics of thermal states following a global quantum quench to a confined phase of a long-range Ising model \cite{LRIM_CONF_1,LRIM_CONF_2,LRIM_CONF_expt}. The post-quench time evolution of domain walls kinks and their Fourier signals have highlighted the intricate interplay between slow-decaying long-range interactions and the emergence of confinement at low temperature. Furthermore, the time dependent fluctuation of domain wall kinks provides a solid evidence of dynamical deconfinement driven by thermally tuned density of excitations.
 
Our study opens up promising avenues for both theoretical exploration and experimental verification. The identification of the initial density as a key parameter influencing the deconfinement transition offers a new direction for experimental investigations in various AMO platforms\cite{DYN_PT_expt_1,DYN_PT_expt_2,LRIM_CONF_expt,LRIM_EXPT1,LRIM_EXPT2,LRIM_EXPT3,LRIM_EXPT4,LRIM_EXPT5,LRIM_EXPT6}. Moreover, the implications of our results extend beyond the regime of condensed matter physics, resonating with the study of confinement in lattice gauge theories \cite{LGT_CONF1,LGT_CONF2,LGT_CONF3,LGT_CONF4,LGT_CONF5,LGT_CONF6,LGT_CONF7}.

\acknowledgements
\noindent \textbf{Acknowledgements}: We acknowledge discussion with Guido Pagano regarding experimental realization of our work in trapped-ion platform. We also acknowledge Alvise Bastianello and Alessio Lerose for fruitful discussion regarding the theoretical aspects of confinement and thermalization. E.T. acknowledge support from the MIUR Programme FARE (MEPH), and from  QUANTERA DYNAMITE PCI2022-132919. M.C. acknowledge support from the PNRR MUR project
PE0000023-NQSTI and by the PRIN 2022 (2022R35ZBF) - PE2 - ``ManyQLowD''.

\appendix

\onecolumngrid

\section{Two kink model} \label{2kink_model}

\begin{figure}[!h]
    \centering
    \includegraphics[width=1.00\linewidth]{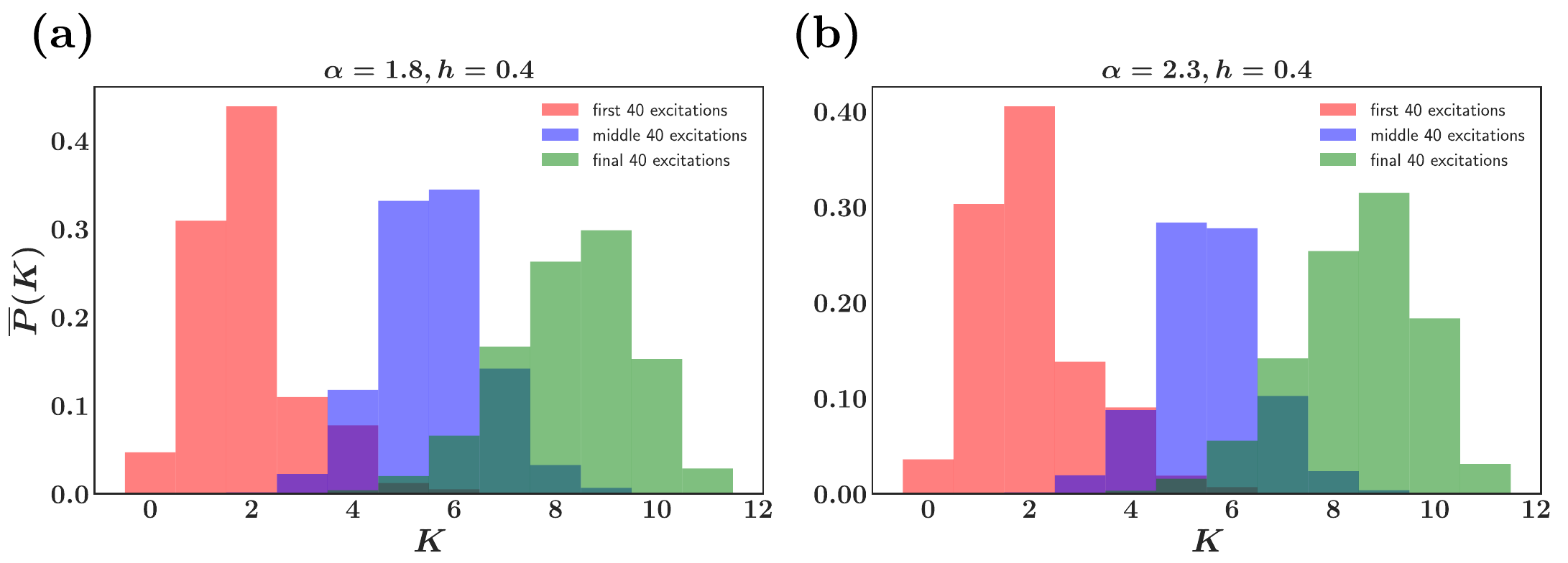}
    \caption{PDF of domain wall kinks averaged over 40 eigenstates at different locations of the full spectrum of two different confining Hamiltonians; $\alpha = 1.8, h = 0.4$ and $\alpha = 2.3, h = 0.4$. The data are for $N=12$ and are computed exactly with ED.}
    \label{fig:kink_conf_ham_spec}
\end{figure}

The initial thermal state can be expanded in the basis of energy eigenstate of the final confined Hamiltonian, $\hat{\rho}_{\beta} = \sum_{m,n} C_{m,n} \ket{m}\bra{n}$. The time-dependent value of the observable can be written as $\langle \hat{k}(t)\rangle = \text{Tr}[\hat{\rho}_{\beta}(t)\hat{k}] = \sum_{m,n} C_{m,n} e^{-it(E_m-E_n)}\langle n |\hat{k}|m\rangle$. If the initial state has a dominant overlap with a few eigenstates of the final Hamiltonian, then $\langle \hat{k}(t)\rangle$ will show oscillations with frequencies corresponding to the energy difference between the eigenstates. The low-temperature state has a dominant overlap with a few low lying eigenstates of the confining Hamiltonian and therefore shows persistent oscillations with sharp frequency peaks; however, initial states with higher temperatures also overlap with higher excited eigenstates, and consequently, no dominant oscillations are observed. Because the observed oscillation frequency is dominated by the low lying energy eigenstates of the final Hamiltonian, it is sufficient to restrict the Hilbert space to a subspace consisting of only low lying eigenstates. In the thermodynamic limit (and with periodic boundary condition), the low lying eigenstates of the long-range confining Hamiltonian are dominantly composed of states with zero- and two-domain wall kinks; however, for a finite system, the low lying eigenstates also have dominant contributions from states with single kinks at the edges along with zero- and two-domain wall kinks. Figure \ref{fig:kink_conf_ham_spec} shows the PDF of the domain wall kinks averaged over 40 eigenstates at different locations of the full spectrum of the post-quench Hamiltonian. We observe that the low lying eigenstates (red bars) are predominantly composed of up to two kinks. 

We use the two kink model \cite{LRIM_CONF_2,LRIM_CONF_expt} to study the low energy excitations as bound quasiparticles in the long-range Ising model. The idea is to project the Hilbert space into a subspace where $n$ spins are clustered together in the sea of up spins, forming up to two domain wall kinks that are free to shift, contract, or expand. The state is defined by two quantum numbers,

\begin{equation}
    \label{eq:state_2kink}
    \ket{j,n} = \ket{...\uparrow\uparrow\downarrow_j\downarrow_{j+1}...\downarrow\downarrow_{j+n-1}\uparrow\uparrow...}
\end{equation}

The long range Ising Hamiltonian can be projected in the in the two kink subspace is $\mathcal{\hat{H}} =  \mathcal{\hat{P}}^{-1}\hat{H}\mathcal{\hat{P}}$, where $\mathcal{\hat{P}}$ is the projection operator,
\begin{equation}
\label{eq:2kinkham}
 \begin{aligned}
    \mathcal{\hat{H}}\ket{j,n} =  V_{\alpha,N}(j,n)\ket{j,n} - h \big[ \ket{j,n+1}+\ket{j,n-1} 
    +\ket{j+1,n-1}+\ket{j-1,n+1}\big]
\end{aligned}  
\end{equation}

where, for a finite system with $N$ spins, $1 \leq j \leq N-1$ and $1 \leq n \leq N-j-1$. The first term is diagonal, $V_{\alpha,N}(j,n)$ is the total potential energy above the ground state (fully polarized), and the second off-diagonal term is the spin flip term, which is responsible for the shift, expansion, and contraction of the domain wall kinks. The projected Hamiltonian \ref{eq:2kinkham} can be diagonalized on the basis $\{\ket{j,n}\}$ and the masses of the mesons can be extracted from the energy spectrum. The matrix we need to diagonalize is

\begin{equation}
    \label{eq:2kink_Ham_Mat}
       \mathcal{H}_{j,n:j',n'} = V_{\alpha,N}(j,n)\delta_{j,j'}\delta_{n,n'} -h\big[\delta_{j,j'}\delta_{n+1,n'}+\delta_{j,j'}\delta_{n-1,n'}+\delta_{j+1,j'}\delta_{n-1,n'}+\delta_{j-1,j'}\delta_{n+1,n'}\big]
\end{equation}

where,

\begin{equation}
    V_{\alpha,N}(j,n) = 2 \sum_{j \leq u \leq j+n-1}\Bigg[ \sum_{1 \leq v \leq j-1} \frac{1}{\abs{v-u}^{\alpha}} + \sum_{j+n \leq v \leq N} \frac{1}{\abs{v-u}^{\alpha}}\Bigg]
\label{eq:2kink_pot}
\end{equation}

is the potential energy of excitation of the two kink states $\ket{j,n}$ above the ground state. This model provides a good description of confinement in a long-range Ising chain in the limit $N \rightarrow \infty$ where the confining potential increases monotonically with distance between the coupled domain walls.\\
    
\section{Single kink model and localization length}\label{sec:1kink_model}

Similar to the two kink model we can define the single kink model by projecting the Hilbert space into a subspace with just one single kink \cite{LRIM_CONF_1}. The quantum state is defined by a single quantum number signifying the position of the single kink,

\begin{equation}
    \label{eq:state_1kink}
    \ket{j,n} = \ket{...\uparrow\uparrow\uparrow_n\downarrow_{n+1}\downarrow\downarrow...}.
\end{equation}

The corresponding Hamiltonian is defined as,

\begin{equation}
    \label{eq:1kink_Ham_Mat}
       \mathcal{H}_{n:n'} = V_{\alpha,N}(n)\delta_{n,n'} -h\big[\delta_{n+1,n'}+\delta_{n-1,n'}\big]
\end{equation}

\begin{figure}[!ht]
    \centering
    \includegraphics[width=1.00\linewidth]{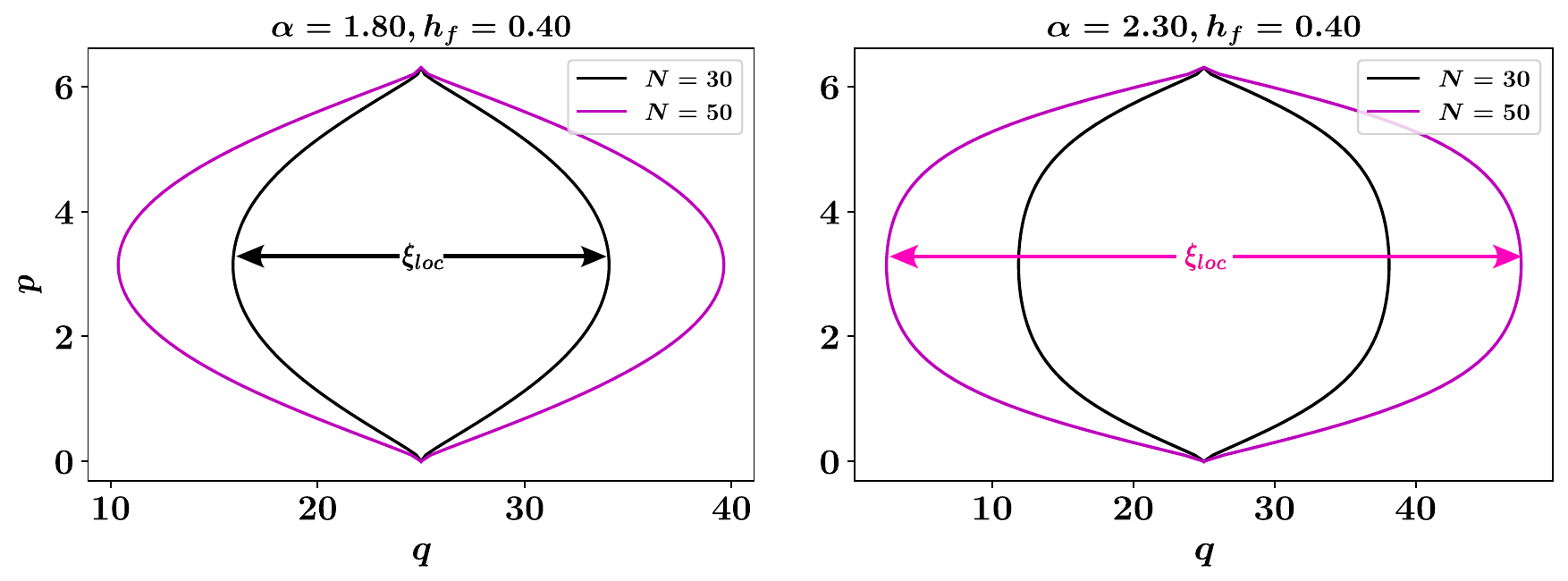}
    \caption{Phase space of the semi-classical Hamiltonian \ref{eq:1kink_semiclassical} over a full period of momentum for two different system parameters. $\xi_{loc}$ is called the localization length and is defined as the maximum space traversed by the single kink originally located at the center of the system.}
    \label{fig:q_k_space}
\end{figure}

where the effective potential is,

\begin{equation}
    V_{\alpha,N}(n) = 2 \sum_{1 \leq i \leq n} \sum_{n+1 \leq j \leq N} \frac{1}{\abs{i-j}^{\alpha}}.
\label{eq:1kink_pot}
\end{equation}

In thermodynamic limit the sums in the potential can be approximated with integrals,

\begin{align}
    V_{\alpha,N}(n) &= 2 \Bigg[ \sum_{r = 1}^{N-n} \frac{1}{r^{\alpha}}+\sum_{r = 2}^{N-n+1} \frac{1}{r^{\alpha}}+....+\sum_{r = n}^{N-1} \frac{1}{r^{\alpha}}\Bigg] \\
    &\approx \Bigg[\int_{1}^{N-n} \frac{dr}{r^{\alpha}}+\int_{2}^{N-n+1} \frac{dr}{r^{\alpha}}+....+\int_{n}^{N-1} \frac{dr}{r^{\alpha}}\Bigg]\\
    & = \frac{1}{(\alpha-1)}\Bigg[\sum_{r=1}^n \frac{1}{r^{\alpha-1}} - \sum_{n'=n}^1 \frac{1}{(N-n')^{\alpha-1}}\Bigg]
\label{eq:1kink_pot_approx1}
\end{align}

On further approximation of the sums we get,

\begin{align}
    V_{\alpha,N}(n) &\approx \frac{2}{(\alpha-1)}\Bigg[\int_{1}^n \frac{dr}{r^{\alpha-1}} - \int_{n}^1 \frac{dn'}{(N-n')^{\alpha-1}}\Bigg]\\
    &= \frac{2}{(\alpha-1)(2-\alpha)} \Bigg[\frac{1}{n^{\alpha-2}} + \frac{1}{(N-n)^{\alpha-2}}-1-\frac{1}{(N-1)^{\alpha-2}} \Bigg]
\label{eq:1kink_pot_approx2}
\end{align}

We can take the classical limit of the Hamiltonian in equation \ref{eq:1kink_Ham_Mat} by defining a phase space $(p,q) \in [0,2\pi]\times [1,N-1]$ and corresponding Hamiltonian,

\begin{equation}
    H^{\text{cls}}_{\alpha,h,N}(p,q) = V_{\alpha,N}(q) - 2h \hspace{0.1cm} \text{cos}(p)
\label{eq:1kink_semiclassical}
\end{equation}

where the function $V_{\alpha,N}(q)$ is defined in the equations \ref{eq:1kink_pot_approx2}. Starting from the most energetic state in a finite chain within the single kink scenario; a single static kink located at the center of the chain, we can calculate the $p \times q$ phase space by solving;

\begin{equation}
    H^{\text{cls}}_{\alpha,N}(0,N/2) =  H^{\text{cls}}_{\alpha,N}(p,q)
\label{eq:phase_sol}
\end{equation}

\begin{figure}[!ht]
    \centering
    \includegraphics[width=1.00\linewidth]{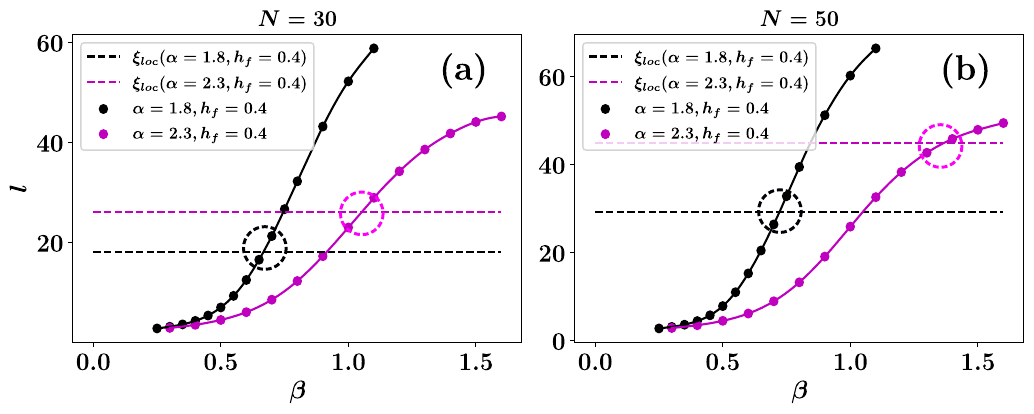}
    \caption{Numerical solution of the equation \ref{eq:extract_T_crit_sup}. The dots are the average kink separation, $l = \frac{N}{\text{Tr}[\hat{\rho}_{\tilde{\beta}}\hat{k}]}$, as a function of $\beta$, the full lines are the cubic interpolation of the data, and the horizontal dashed lines are $\xi_{loc}$ for the corresponding parameters. The dashed circles highlights the point of solution.}
    \label{fig:TC_extract}
\end{figure}

In figure \ref{fig:q_k_space} we plot the classical phase space by solving \ref{eq:phase_sol} over a full periodicity of the momentum. We observe that the kink travels farthest from its original position at $p = \pi$. We define this distance as the localization length of the kink, $\xi_{loc}$. Localization length separates two different dynamical regimes: when the average separation between the kinks is larger than $\xi_{loc}$ the kinks exhibit uncorrelated Bloch oscillations strictly localized at their site of origin, as the average kink separation between kinks becomes comparable to and smaller than $\xi_{loc}$ the kinks disperse and delocalize. The critical temperature corresponding to this transition can be extracted by numerically solving the following equation for $\beta$;

\begin{equation}
    l = \frac{N}{\overline{\langle \hat{k}_{\beta}(t) \rangle}} =  \xi_{loc},
\label{eq:extract_T_crit_sup}
\end{equation}

where $\overline{\langle \hat{k}_{\beta}(t) \rangle}$ is the time average of $\text{Tr}[\hat{\rho}_{\beta}(t)\hat{k}]$. The post-quench behavior of $\langle \hat{k}_{\beta}(t) \rangle$ suggests that we can replace $\overline{\langle \hat{k}_{\beta}(t) \rangle} $ with the expected thermal kink density $ \text{Tr}[\hat{\rho}_{\tilde{\beta}}\hat{k}]$.

\section{Simulation details}\label{sec:sim_det}

\subsection{Simulation of finite temperature states}\label{subsec:fin_temp}

The finite temperature states are simulated by the method of purification \cite{fintem1,fintem2}. We begin at infinite temperature, $\beta = 0$, where the state is maximally mixed and can be written as the tensor product of local identities $\hat{\rho}_{0} = \bigotimes_{i=1}^{N}\mathbf{1}^{\sigma_i,\Tilde{\sigma}_i} = \mathbbm{1}$, where $\mathbf{1}^{\sigma_i,\Tilde{\sigma}_i} = [\delta_{\sigma_i,\Tilde{\sigma}_i}]_{d \times d}$ and $d$ is the dimension of the physical space. The density operator for any non-zero $\beta$ is 

\begin{subequations}
\label{eq:fin_tem}
\begin{align}
\hat{\rho}_{\beta} \propto  e^{-\beta \hat{H}} &= e^{-\frac{\beta}{2}\hat{H}} \mathbbm{1} e^{-\frac{\beta }{2}\hat{H}}\\
&\propto e^{-\frac{\beta}{2}\hat{H}} \hat{\rho}_{0} e^{-\frac{\beta}{2}\hat{H}}
\end{align}
\end{subequations}

\begin{figure}[!ht]
    \centering
    \includegraphics[width=1.0\linewidth]{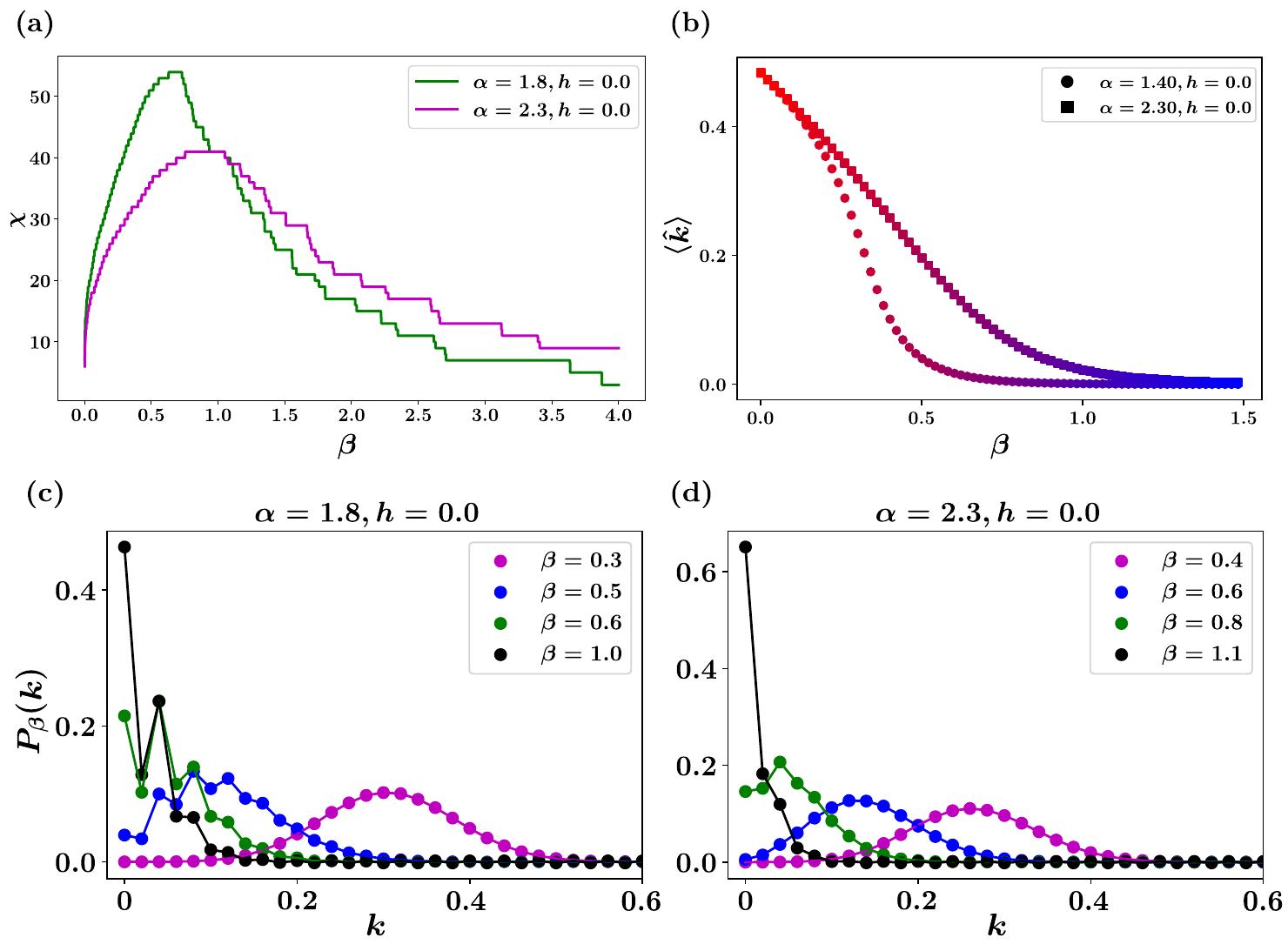}
    \caption{(a) Bond dimension corresponding to the central site as a function of inverse temperature during the imaginary time evolution. All singular values smaller than $10^{-8}$ are discarded during truncation. (b) Average kink density $\langle \hat{k} \rangle $ as a function of inverse temperature $\beta$ for three different parameters and system size $N=51$. The figure illustrates a thermal phase transition from low temperature ferromagnetic regime (in blue) to high temperature paramagnetic regime (in red) based on the kink density. Notice that at $\beta = 0$ kink density is 0.5 for all parameter values which is the expected value for a maximally mixed state. (c)-(d) PDF of domain wall kinks in the initial thermal state at four representative temperatures with $\alpha = 1.8$ and $\alpha = 2.3$ respectively.} 
    \label{fig:kink_den_vs_temp}
\end{figure}

We keep the density operator in locally purified form $\hat{\rho} = \mathbb{X}\mathbb{X}^{\dagger}$ at each stage where $\mathbb{X}$ is represented as tensor

\begin{equation}
    \label{eq:MPDO_half}
    \mathbb{X}^{\sigma_1,\sigma_2,...\sigma_i,...,\sigma_N}_{k_1,k_2,...,k_i,...,k_N} = A^{\sigma_1,k_1}_{m_0,m_1}A^{\sigma_2,k_2}_{m_1,m_2}...A^{\sigma_i,k_i}_{m_{i-1},m_i}...A^{\sigma_N,k_N}_{m_{N-1},m_N}
\end{equation}

where $s_i = d$, $k_i = d$ are the physical index and the Kraus index are are fixed through out and $1 \leq m_i \leq \chi_{max}$ is the bond index. The density operator can now be purified to a given $\beta$ in trotterized steps

\begin{subequations}
    \label{eq:purify}
    \begin{align}
    \hat{\rho}_{\beta+d\beta} &= e^{-\frac{d\beta}{2}\hat{H}} \hat{\rho}_{\beta} e^{-\frac{d\beta }{2}\hat{H}}\\
    &= e^{-\frac{d\beta}{2}\hat{H}} \mathbb{X}_{\beta}\mathbb{X}_{\beta}^{\dagger} e^{-\frac{d\beta}{2}\hat{H}}\\
    &= e^{-\frac{d\beta}{2}\hat{H}} \mathbb{X}_{\beta}[e^{-\frac{d\beta}{2}\hat{H}} \mathbb{X}_{\beta}]^{\dagger}
    \end{align}
\end{subequations}

The simulation of Equation (\ref{eq:purify}) can be achieved through an imaginary time Time-Dependent Variational Principle (TDVP) \cite{TDVP_one,TDVP_two} by employing the transformation $-idt \rightarrow -d\beta$, while rigorously maintaining the locally purified form. It is sufficient to simulate $\mathbb{X}_{\beta+d\beta} = e^{-\frac{d\beta}{2}\hat{H}} \mathbb{X}_{\beta}$; the other half is a trivial conjugate. We employ a two-site TDVP algorithm with a time step of $d\beta = 0.001$. The initial state, denoted by $\hat{\rho_0}$, is maximally mixed and has a small bond dimension of two. Notably, unlike real-time evolution, the bond dimension does not exhibit excessive growth during imaginary time evolution.

In Figure \ref{fig:kink_den_vs_temp} (a), the bond dimension corresponding to the central site is plotted against $\beta$. Truncation involves discarding all singular values smaller than $10^{-8}$. The peaks in the plot indicate critical regions where the area law is invalid \cite{area_law_1,area_law_3,area_law_5,area_law_6,area_law_7}.
In Figure \ref{fig:kink_den_vs_temp} (b), the average kink density $\langle \hat{k} \rangle$ is plotted as a function of the inverse temperature $\beta$ for different $\alpha$ and $h = 0.0$. At $\beta = 0$, the state is maximally mixed and $\langle \hat{k} \rangle = 0.5$ holds for all parameters. As $\beta$ increases, there is a monotonic reduction in $\langle \hat{k} \rangle$, indicating a thermal transition into the ferromagnetic phase, where $\langle \hat{k} \rangle = 0$. It is important to note that the thermal phase transition observed in Figure \ref{fig:kink_den_vs_temp} is robust only for $\alpha \leq 2$ in the long-range Ising model. The transition observed for $\alpha = 2.3$ will crossover to $\beta = \infty$ as $N \rightarrow \infty$. This implies the thermal phase transition at $T = 0$ or equivalently absence of ferromagnetic order in LRIM for $\alpha > 2$. Figure \ref{fig:kink_den_vs_temp} (c)–(d) illustrates the corresponding PDF of the domain-wall kink density for the initial states at four representative temperatures. At high temperatures (low $\beta$), the kink distribution appears broad, whereas it peaks sharply at zero at low temperatures.

\subsection{Real time evolution of thermal state}

\begin{figure}[!t]
    \centering
    \includegraphics[width=1.00\linewidth]{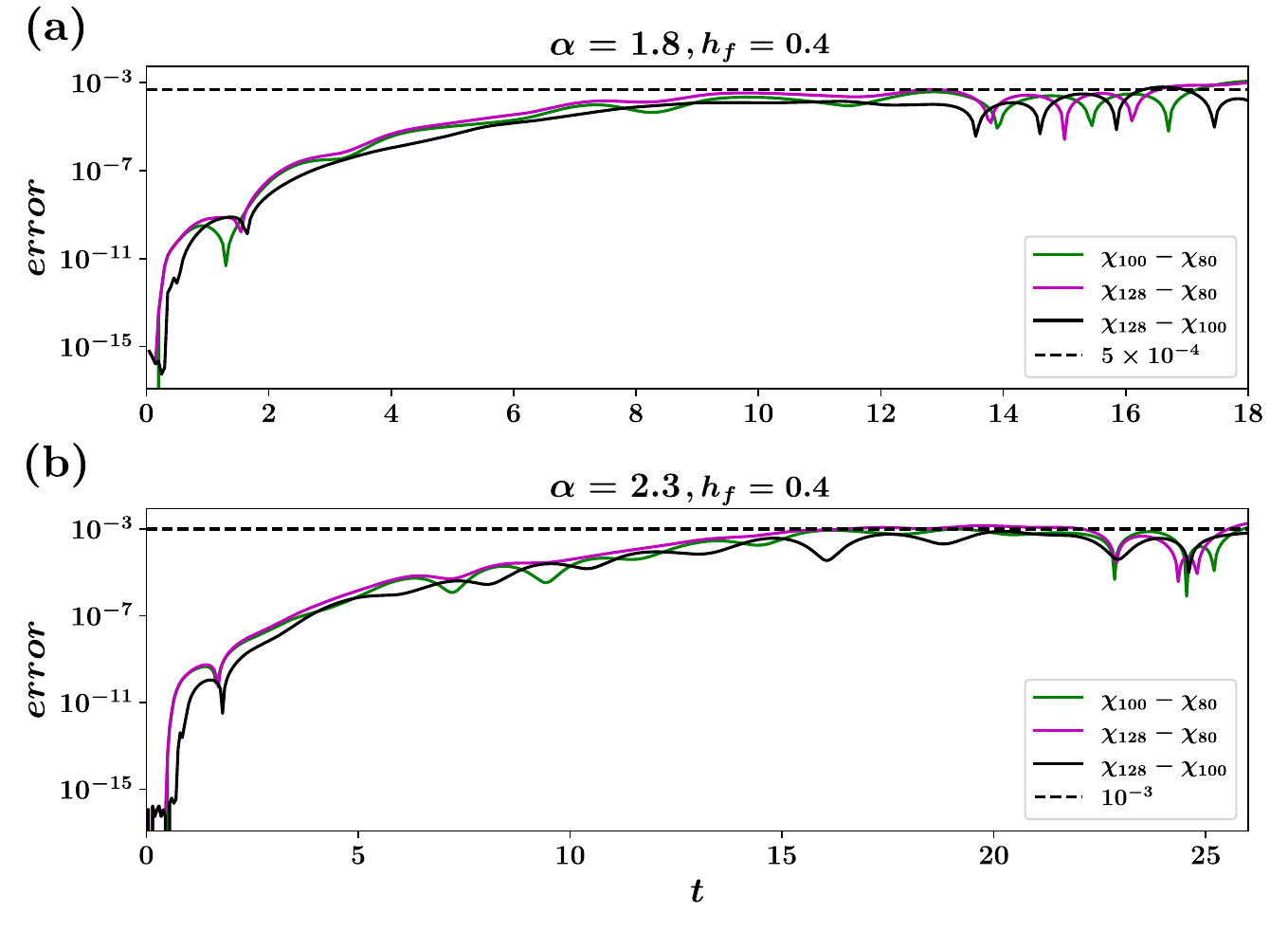}
    \caption{Convergence of the TDVP data for DT(t) with increasing bond dimensions, $\chi = 80, 100, 128$, for $\alpha = 1.8, h_f = 0.4$ (a) and $\alpha = 2.3, h_f = 0.4$ (b). The black dashed line is for visual guidance.}
    \label{fig:rel_err}
\end{figure}

The thermal density operator can be evolved in real time while keeping the locally purified form intact,

\begin{subequations}
    \label{eq:real_time_rho}
    \begin{align}
    \hat{\rho}_{\beta}(t+dt) &= e^{-i dt \hat{H}}\hat{\rho}_{\beta}(t)e^{i dt \hat{H}}\\
    &= e^{-i dt \hat{H}}\mathbb{X}_{\beta}(t) \mathbb{X}^{\dagger}_{\beta}(t) e^{i dt \hat{H}}\\
    &= e^{-i dt \hat{H}}\mathbb{X}_{\beta}(t) \Big[ e^{-i dt \hat{H}}\mathbb{X}_{\beta}(t) \Big]^{\dagger}
    \end{align}
\end{subequations}

We employ a two-site Time-Dependent Variational Principle (TDVP) algorithm with a time step of $dt = 0.05$ for real-time evolution. Similar to the imaginary time evolution, it is sufficient to simulate $\mathbb{X}_{\beta}(t+dt) = e^{-i dt \hat{H}}\mathbb{X}_{\beta}(t)$ with the other half being its trivial conjugate. To assess the convergence of the TDVP data, we calculate the relative error in kink density using three increasing bond dimensions ($\chi_{\text{max}} = 80, 100, 128$) at $\beta = 0.3$, as illustrated in Figure \ref{fig:rel_err}. Relative errors consistently remain below $O(10^{-3})$.

The explanation for choosing $\beta = 0.3$ to test the convergence of all TDVP data is straightforward. Increasing the temperature facilitates correlation spreading through the system during real-time evolution, necessitating a higher bond dimension to capture the dynamics effectively.  Thus, testing the convergence of errors for the worst-case scenario, where $\beta = 0.3$  is sufficient for our study. The results discussed in the main text are based on a maximum bond dimension of $\chi_{\text{max}} = 128$.

\subsection{Full counting statistics of kink density}\label{subsec:fcs_kink}

The full counting statistics of a generic quantum mechanical operator $\hat{O}$ over a density matrix $\hat{\rho}$ can be calculated as;

\begin{equation}\label{eq:PDF_RHO_1}
    P(o) = \text{Tr}[\hat{\rho}\delta(\hat{O}-o)]
\end{equation}

\begin{figure}[!ht]
    \centering
    \includegraphics[width=1.00\linewidth]{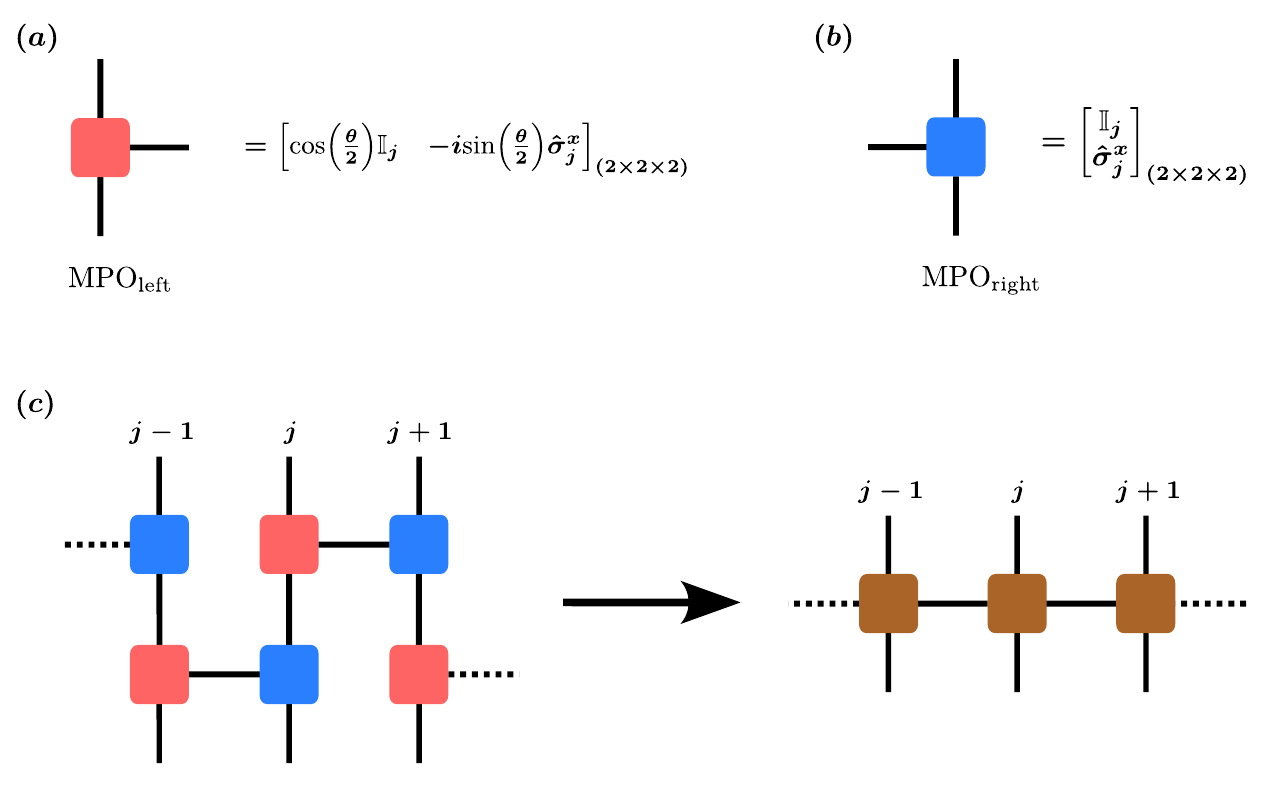}
    \caption{Left (red) and right (blue) MPO at site $j$, (a) and (b) respectively. Contracting left and right MPO at each site to build a four legged MPO (black)  at each site, (c).}
    \label{fig:GF_kink}
\end{figure}

which can be Fourier transformed to an integral

\begin{equation}\label{eq:PDF_RHO_2}
    P(o) = \int_{-\infty}^{\infty} \frac{d\theta}{2\pi} e^{-i\theta o} \text{Tr}\big[\hat{\rho} e^{i\theta \hat{O}}\big].
\end{equation}

In our specific case the operator is the kink density $\hat{k}$ defined as,

\begin{equation}\label{eq:supp_kink_den}
    \hat{k} = \frac{1}{N}\sum_{j=1}^{N-1} 1-\hat{\sigma}^x_j \hat{\sigma}^x_{j+1}
\end{equation}

replacing this in equation \ref{eq:PDF_RHO_2} (without the extensive $1/N$) gives us,

\begin{equation}\label{eq:PDF_RHO_3}
    P(k) = \int_{-\infty}^{\infty} \frac{d\theta}{2\pi} e^{-i\theta \Big[k-\frac{N-1}{2}\Big]} \text{Tr}\Bigg[\hat{\rho} \prod_{j=1}^{N-1} e^{-i\theta \frac{\hat{\sigma}^x_j\hat{\sigma}^x_{j+1}}{2} }\Bigg].
\end{equation}

The problem boils down to calculating the trace of the product of two site exponential operator. We can expand the two site exponential operator by Taylor series and rearrange to break it down into a product of two single site operators acting on site $i$ and $i+1$ respectively, 

\begin{equation}\label{eq:2sitebreakdown}
    e^{-i\theta \frac{\hat{\sigma}^x_j\hat{\sigma}^x_{j+1}}{2}} = \text{cos}\Big(\frac{\theta}{2}\Big) - i\hspace{0.1cm} \text{sin}\Big(\frac{\theta}{2}\Big)\hat{\sigma}^x_j\hat{\sigma}^x_{j+1}  = \begin{bmatrix}
    \text{cos}\Big(\frac{\theta}{2}\Big) \mathbbm{I}_j & -i\hspace{0.1cm} \text{sin}\Big(\frac{\theta}{2}\Big)\hat{\sigma}^x_j
    \end{bmatrix}
    \begin{bmatrix}
        \mathbbm{I}_{j+1}\\
        \hat{\sigma}^x_{j+1}
    \end{bmatrix}
\end{equation}

Equation \ref{eq:2sitebreakdown} suggests that that the integral in equation \ref{eq:PDF_RHO_3} has a periodicity of $2\pi$ so the integral can be restricted to $\theta \in [-\pi,\pi]$. We reshape and redefine the operators acting on site $i$ and $i+1$ in equation \ref{eq:2sitebreakdown} as $\text{MPO}_{\text{left}}$ and $\text{MPO}_{\text{right}}$ respectively (see (a) and (b) in figure \ref{fig:GF_kink}). Each site now has two single site tensors that can be compressed together into a four legged MPO (see (c) in figure \ref{fig:GF_kink}) which can act on a matrix product density operator in a straight forward manner.

\bibliography{bibliography}

\begin{thebibliography}{83}%
\makeatletter
\providecommand \@ifxundefined [1]{%
 \@ifx{#1\undefined}
}%
\providecommand \@ifnum [1]{%
 \ifnum #1\expandafter \@firstoftwo
 \else \expandafter \@secondoftwo
 \fi
}%
\providecommand \@ifx [1]{%
 \ifx #1\expandafter \@firstoftwo
 \else \expandafter \@secondoftwo
 \fi
}%
\providecommand \natexlab [1]{#1}%
\providecommand \enquote  [1]{``#1''}%
\providecommand \bibnamefont  [1]{#1}%
\providecommand \bibfnamefont [1]{#1}%
\providecommand \citenamefont [1]{#1}%
\providecommand \href@noop [0]{\@secondoftwo}%
\providecommand \href [0]{\begingroup \@sanitize@url \@href}%
\providecommand \@href[1]{\@@startlink{#1}\@@href}%
\providecommand \@@href[1]{\endgroup#1\@@endlink}%
\providecommand \@sanitize@url [0]{\catcode `\\12\catcode `\$12\catcode `\&12\catcode `\#12\catcode `\^12\catcode `\_12\catcode `\%12\relax}%
\providecommand \@@startlink[1]{}%
\providecommand \@@endlink[0]{}%
\providecommand \url  [0]{\begingroup\@sanitize@url \@url }%
\providecommand \@url [1]{\endgroup\@href {#1}{\urlprefix }}%
\providecommand \urlprefix  [0]{URL }%
\providecommand \Eprint [0]{\href }%
\providecommand \doibase [0]{https://doi.org/}%
\providecommand \selectlanguage [0]{\@gobble}%
\providecommand \bibinfo  [0]{\@secondoftwo}%
\providecommand \bibfield  [0]{\@secondoftwo}%
\providecommand \translation [1]{[#1]}%
\providecommand \BibitemOpen [0]{}%
\providecommand \bibitemStop [0]{}%
\providecommand \bibitemNoStop [0]{.\EOS\space}%
\providecommand \EOS [0]{\spacefactor3000\relax}%
\providecommand \BibitemShut  [1]{\csname bibitem#1\endcsname}%
\let\auto@bib@innerbib\@empty
\bibitem [{\citenamefont {Coleman}(2015)}]{coleman2015introduction}%
  \BibitemOpen
  \bibfield  {author} {\bibinfo {author} {\bibfnamefont {P.}~\bibnamefont {Coleman}},\ }\href@noop {} {\emph {\bibinfo {title} {Introduction to many-body physics}}}\ (\bibinfo  {publisher} {Cambridge University Press},\ \bibinfo {year} {2015})\BibitemShut {NoStop}%
\bibitem [{\citenamefont {Greiner}\ \emph {et~al.}(2007)\citenamefont {Greiner}, \citenamefont {Schramm},\ and\ \citenamefont {Stein}}]{greiner2007quantum}%
  \BibitemOpen
  \bibfield  {author} {\bibinfo {author} {\bibfnamefont {W.}~\bibnamefont {Greiner}}, \bibinfo {author} {\bibfnamefont {S.}~\bibnamefont {Schramm}},\ and\ \bibinfo {author} {\bibfnamefont {E.}~\bibnamefont {Stein}},\ }\href@noop {} {\emph {\bibinfo {title} {Quantum chromodynamics}}}\ (\bibinfo  {publisher} {Springer Science \& Business Media},\ \bibinfo {year} {2007})\BibitemShut {NoStop}%
\bibitem [{\citenamefont {Busza}\ \emph {et~al.}(2018)\citenamefont {Busza}, \citenamefont {Rajagopal},\ and\ \citenamefont {Van Der~Schee}}]{busza2018heavy}%
  \BibitemOpen
  \bibfield  {author} {\bibinfo {author} {\bibfnamefont {W.}~\bibnamefont {Busza}}, \bibinfo {author} {\bibfnamefont {K.}~\bibnamefont {Rajagopal}},\ and\ \bibinfo {author} {\bibfnamefont {W.}~\bibnamefont {Van Der~Schee}},\ }\bibfield  {title} {\bibinfo {title} {Heavy ion collisions: the big picture and the big questions},\ }\href@noop {} {\bibfield  {journal} {\bibinfo  {journal} {Annual Review of Nuclear and Particle Science}\ }\textbf {\bibinfo {volume} {68}},\ \bibinfo {pages} {339} (\bibinfo {year} {2018})}\BibitemShut {NoStop}%
\bibitem [{\citenamefont {Berges}\ \emph {et~al.}(2021)\citenamefont {Berges}, \citenamefont {Heller}, \citenamefont {Mazeliauskas},\ and\ \citenamefont {Venugopalan}}]{berges2021qcd}%
  \BibitemOpen
  \bibfield  {author} {\bibinfo {author} {\bibfnamefont {J.}~\bibnamefont {Berges}}, \bibinfo {author} {\bibfnamefont {M.~P.}\ \bibnamefont {Heller}}, \bibinfo {author} {\bibfnamefont {A.}~\bibnamefont {Mazeliauskas}},\ and\ \bibinfo {author} {\bibfnamefont {R.}~\bibnamefont {Venugopalan}},\ }\bibfield  {title} {\bibinfo {title} {Qcd thermalization: Ab initio approaches and interdisciplinary connections},\ }\href@noop {} {\bibfield  {journal} {\bibinfo  {journal} {Reviews of Modern Physics}\ }\textbf {\bibinfo {volume} {93}},\ \bibinfo {pages} {035003} (\bibinfo {year} {2021})}\BibitemShut {NoStop}%
\bibitem [{\citenamefont {Rothkopf}(2020)}]{rothkopf2020heavy}%
  \BibitemOpen
  \bibfield  {author} {\bibinfo {author} {\bibfnamefont {A.}~\bibnamefont {Rothkopf}},\ }\bibfield  {title} {\bibinfo {title} {Heavy quarkonium in extreme conditions},\ }\href@noop {} {\bibfield  {journal} {\bibinfo  {journal} {Physics Reports}\ }\textbf {\bibinfo {volume} {858}},\ \bibinfo {pages} {1} (\bibinfo {year} {2020})}\BibitemShut {NoStop}%
\bibitem [{\citenamefont {Chanda}\ \emph {et~al.}(2020)\citenamefont {Chanda}, \citenamefont {Zakrzewski}, \citenamefont {Lewenstein},\ and\ \citenamefont {Tagliacozzo}}]{LGT_CONF1}%
  \BibitemOpen
  \bibfield  {author} {\bibinfo {author} {\bibfnamefont {T.}~\bibnamefont {Chanda}}, \bibinfo {author} {\bibfnamefont {J.}~\bibnamefont {Zakrzewski}}, \bibinfo {author} {\bibfnamefont {M.}~\bibnamefont {Lewenstein}},\ and\ \bibinfo {author} {\bibfnamefont {L.}~\bibnamefont {Tagliacozzo}},\ }\bibfield  {title} {\bibinfo {title} {Confinement and lack of thermalization after quenches in the bosonic schwinger model},\ }\href {https://doi.org/10.1103/PhysRevLett.124.180602} {\bibfield  {journal} {\bibinfo  {journal} {Phys. Rev. Lett.}\ }\textbf {\bibinfo {volume} {124}},\ \bibinfo {pages} {180602} (\bibinfo {year} {2020})}\BibitemShut {NoStop}%
\bibitem [{\citenamefont {Sedgewick}\ \emph {et~al.}(2002)\citenamefont {Sedgewick}, \citenamefont {Scalapino},\ and\ \citenamefont {Sugar}}]{LGT_CONF2}%
  \BibitemOpen
  \bibfield  {author} {\bibinfo {author} {\bibfnamefont {R.~D.}\ \bibnamefont {Sedgewick}}, \bibinfo {author} {\bibfnamefont {D.~J.}\ \bibnamefont {Scalapino}},\ and\ \bibinfo {author} {\bibfnamefont {R.~L.}\ \bibnamefont {Sugar}},\ }\bibfield  {title} {\bibinfo {title} {Fractionalized phase in an $\mathrm{XY}--{Z}_{2}$ gauge model},\ }\href {https://doi.org/10.1103/PhysRevB.65.054508} {\bibfield  {journal} {\bibinfo  {journal} {Phys. Rev. B}\ }\textbf {\bibinfo {volume} {65}},\ \bibinfo {pages} {054508} (\bibinfo {year} {2002})}\BibitemShut {NoStop}%
\bibitem [{\citenamefont {Kühn}\ \emph {et~al.}(2015)\citenamefont {Kühn}, \citenamefont {Zohar}, \citenamefont {Cirac},\ and\ \citenamefont {Ba{\~{n}}uls}}]{LGT_CONF3}%
  \BibitemOpen
  \bibfield  {author} {\bibinfo {author} {\bibfnamefont {S.}~\bibnamefont {Kühn}}, \bibinfo {author} {\bibfnamefont {E.}~\bibnamefont {Zohar}}, \bibinfo {author} {\bibfnamefont {J.~I.}\ \bibnamefont {Cirac}},\ and\ \bibinfo {author} {\bibfnamefont {M.~C.}\ \bibnamefont {Ba{\~{n}}uls}},\ }\bibfield  {title} {\bibinfo {title} {Non-abelian string breaking phenomena with matrix product states},\ }\bibfield  {journal} {\bibinfo  {journal} {Journal of High Energy Physics}\ }\textbf {\bibinfo {volume} {2015}},\ \href {https://doi.org/10.1007/jhep07(2015)130} {10.1007/jhep07(2015)130} (\bibinfo {year} {2015})\BibitemShut {NoStop}%
\bibitem [{\citenamefont {Pichler}\ \emph {et~al.}(2016)\citenamefont {Pichler}, \citenamefont {Dalmonte}, \citenamefont {Rico}, \citenamefont {Zoller},\ and\ \citenamefont {Montangero}}]{LGT_CONF4}%
  \BibitemOpen
  \bibfield  {author} {\bibinfo {author} {\bibfnamefont {T.}~\bibnamefont {Pichler}}, \bibinfo {author} {\bibfnamefont {M.}~\bibnamefont {Dalmonte}}, \bibinfo {author} {\bibfnamefont {E.}~\bibnamefont {Rico}}, \bibinfo {author} {\bibfnamefont {P.}~\bibnamefont {Zoller}},\ and\ \bibinfo {author} {\bibfnamefont {S.}~\bibnamefont {Montangero}},\ }\bibfield  {title} {\bibinfo {title} {Real-time dynamics in u(1) lattice gauge theories with tensor networks},\ }\href {https://doi.org/10.1103/PhysRevX.6.011023} {\bibfield  {journal} {\bibinfo  {journal} {Phys. Rev. X}\ }\textbf {\bibinfo {volume} {6}},\ \bibinfo {pages} {011023} (\bibinfo {year} {2016})}\BibitemShut {NoStop}%
\bibitem [{\citenamefont {Zohar}\ \emph {et~al.}(2012)\citenamefont {Zohar}, \citenamefont {Cirac},\ and\ \citenamefont {Reznik}}]{LGT_CONF5}%
  \BibitemOpen
  \bibfield  {author} {\bibinfo {author} {\bibfnamefont {E.}~\bibnamefont {Zohar}}, \bibinfo {author} {\bibfnamefont {J.~I.}\ \bibnamefont {Cirac}},\ and\ \bibinfo {author} {\bibfnamefont {B.}~\bibnamefont {Reznik}},\ }\bibfield  {title} {\bibinfo {title} {Simulating compact quantum electrodynamics with ultracold atoms: Probing confinement and nonperturbative effects},\ }\href {https://doi.org/10.1103/PhysRevLett.109.125302} {\bibfield  {journal} {\bibinfo  {journal} {Phys. Rev. Lett.}\ }\textbf {\bibinfo {volume} {109}},\ \bibinfo {pages} {125302} (\bibinfo {year} {2012})}\BibitemShut {NoStop}%
\bibitem [{\citenamefont {Barros}\ \emph {et~al.}(2019)\citenamefont {Barros}, \citenamefont {Dalmonte},\ and\ \citenamefont {Trombettoni}}]{LGT_CONF6}%
  \BibitemOpen
  \bibfield  {author} {\bibinfo {author} {\bibfnamefont {J.~C.~P.}\ \bibnamefont {Barros}}, \bibinfo {author} {\bibfnamefont {M.}~\bibnamefont {Dalmonte}},\ and\ \bibinfo {author} {\bibfnamefont {A.}~\bibnamefont {Trombettoni}},\ }\bibfield  {title} {\bibinfo {title} {String tension and robustness of confinement properties in the schwinger-thirring model},\ }\href {https://doi.org/10.1103/PhysRevD.100.036009} {\bibfield  {journal} {\bibinfo  {journal} {Phys. Rev. D}\ }\textbf {\bibinfo {volume} {100}},\ \bibinfo {pages} {036009} (\bibinfo {year} {2019})}\BibitemShut {NoStop}%
\bibitem [{\citenamefont {Magnifico}\ \emph {et~al.}(2020)\citenamefont {Magnifico}, \citenamefont {Dalmonte}, \citenamefont {Facchi}, \citenamefont {Pascazio}, \citenamefont {Pepe},\ and\ \citenamefont {Ercolessi}}]{LGT_CONF7}%
  \BibitemOpen
  \bibfield  {author} {\bibinfo {author} {\bibfnamefont {G.}~\bibnamefont {Magnifico}}, \bibinfo {author} {\bibfnamefont {M.}~\bibnamefont {Dalmonte}}, \bibinfo {author} {\bibfnamefont {P.}~\bibnamefont {Facchi}}, \bibinfo {author} {\bibfnamefont {S.}~\bibnamefont {Pascazio}}, \bibinfo {author} {\bibfnamefont {F.~V.}\ \bibnamefont {Pepe}},\ and\ \bibinfo {author} {\bibfnamefont {E.}~\bibnamefont {Ercolessi}},\ }\bibfield  {title} {\bibinfo {title} {Real {T}ime {D}ynamics and {C}onfinement in the {$\mathbb{Z}_{n}$} {S}chwinger-{W}eyl lattice model for 1+1 {QED}},\ }\href {https://doi.org/10.22331/q-2020-06-15-281} {\bibfield  {journal} {\bibinfo  {journal} {{Quantum}}\ }\textbf {\bibinfo {volume} {4}},\ \bibinfo {pages} {281} (\bibinfo {year} {2020})}\BibitemShut {NoStop}%
\bibitem [{\citenamefont {Zhang}\ \emph {et~al.}(2023)\citenamefont {Zhang}, \citenamefont {Liu}, \citenamefont {Cheng}, \citenamefont {He}, \citenamefont {Wang}, \citenamefont {Wang}, \citenamefont {Zhu}, \citenamefont {Su}, \citenamefont {Zhou}, \citenamefont {Zheng}, \citenamefont {Sun}, \citenamefont {Yang}, \citenamefont {Hauke}, \citenamefont {Zheng}, \citenamefont {Halimeh}, \citenamefont {Yuan},\ and\ \citenamefont {Pan}}]{zhang2023observation}%
  \BibitemOpen
  \bibfield  {author} {\bibinfo {author} {\bibfnamefont {W.-Y.}\ \bibnamefont {Zhang}}, \bibinfo {author} {\bibfnamefont {Y.}~\bibnamefont {Liu}}, \bibinfo {author} {\bibfnamefont {Y.}~\bibnamefont {Cheng}}, \bibinfo {author} {\bibfnamefont {M.-G.}\ \bibnamefont {He}}, \bibinfo {author} {\bibfnamefont {H.-Y.}\ \bibnamefont {Wang}}, \bibinfo {author} {\bibfnamefont {T.-Y.}\ \bibnamefont {Wang}}, \bibinfo {author} {\bibfnamefont {Z.-H.}\ \bibnamefont {Zhu}}, \bibinfo {author} {\bibfnamefont {G.-X.}\ \bibnamefont {Su}}, \bibinfo {author} {\bibfnamefont {Z.-Y.}\ \bibnamefont {Zhou}}, \bibinfo {author} {\bibfnamefont {Y.-G.}\ \bibnamefont {Zheng}}, \bibinfo {author} {\bibfnamefont {H.}~\bibnamefont {Sun}}, \bibinfo {author} {\bibfnamefont {B.}~\bibnamefont {Yang}}, \bibinfo {author} {\bibfnamefont {P.}~\bibnamefont {Hauke}}, \bibinfo {author} {\bibfnamefont {W.}~\bibnamefont {Zheng}}, \bibinfo {author} {\bibfnamefont {J.~C.}\ \bibnamefont {Halimeh}}, \bibinfo {author} {\bibfnamefont {Z.-S.}\ \bibnamefont {Yuan}},\
  and\ \bibinfo {author} {\bibfnamefont {J.-W.}\ \bibnamefont {Pan}},\ }\href {https://arxiv.org/abs/2306.11794} {\bibinfo {title} {Observation of microscopic confinement dynamics by a tunable topological $\theta$-angle}} (\bibinfo {year} {2023}),\ \Eprint {https://arxiv.org/abs/2306.11794} {arXiv:2306.11794 [cond-mat.quant-gas]} \BibitemShut {NoStop}%
\bibitem [{\citenamefont {Halimeh}\ \emph {et~al.}(2022)\citenamefont {Halimeh}, \citenamefont {McCulloch}, \citenamefont {Yang},\ and\ \citenamefont {Hauke}}]{PRXQuantum.3.040316}%
  \BibitemOpen
  \bibfield  {author} {\bibinfo {author} {\bibfnamefont {J.~C.}\ \bibnamefont {Halimeh}}, \bibinfo {author} {\bibfnamefont {I.~P.}\ \bibnamefont {McCulloch}}, \bibinfo {author} {\bibfnamefont {B.}~\bibnamefont {Yang}},\ and\ \bibinfo {author} {\bibfnamefont {P.}~\bibnamefont {Hauke}},\ }\bibfield  {title} {\bibinfo {title} {Tuning the topological $\ensuremath{\theta}$-angle in cold-atom quantum simulators of gauge theories},\ }\href {https://doi.org/10.1103/PRXQuantum.3.040316} {\bibfield  {journal} {\bibinfo  {journal} {PRX Quantum}\ }\textbf {\bibinfo {volume} {3}},\ \bibinfo {pages} {040316} (\bibinfo {year} {2022})}\BibitemShut {NoStop}%
\bibitem [{\citenamefont {Lerose}\ \emph {et~al.}(2019)\citenamefont {Lerose}, \citenamefont {\ifmmode \check{Z}\else \v{Z}\fi{}unkovi\ifmmode~\check{c}\else \v{c}\fi{}}, \citenamefont {Silva},\ and\ \citenamefont {Gambassi}}]{LRIM_CONF_1}%
  \BibitemOpen
  \bibfield  {author} {\bibinfo {author} {\bibfnamefont {A.}~\bibnamefont {Lerose}}, \bibinfo {author} {\bibfnamefont {B.}~\bibnamefont {\ifmmode \check{Z}\else \v{Z}\fi{}unkovi\ifmmode~\check{c}\else \v{c}\fi{}}}, \bibinfo {author} {\bibfnamefont {A.}~\bibnamefont {Silva}},\ and\ \bibinfo {author} {\bibfnamefont {A.}~\bibnamefont {Gambassi}},\ }\bibfield  {title} {\bibinfo {title} {Quasilocalized excitations induced by long-range interactions in translationally invariant quantum spin chains},\ }\href {https://doi.org/10.1103/PhysRevB.99.121112} {\bibfield  {journal} {\bibinfo  {journal} {Phys. Rev. B}\ }\textbf {\bibinfo {volume} {99}},\ \bibinfo {pages} {121112} (\bibinfo {year} {2019})}\BibitemShut {NoStop}%
\bibitem [{\citenamefont {Liu}\ \emph {et~al.}(2019)\citenamefont {Liu}, \citenamefont {Lundgren}, \citenamefont {Titum}, \citenamefont {Pagano}, \citenamefont {Zhang}, \citenamefont {Monroe},\ and\ \citenamefont {Gorshkov}}]{LRIM_CONF_2}%
  \BibitemOpen
  \bibfield  {author} {\bibinfo {author} {\bibfnamefont {F.}~\bibnamefont {Liu}}, \bibinfo {author} {\bibfnamefont {R.}~\bibnamefont {Lundgren}}, \bibinfo {author} {\bibfnamefont {P.}~\bibnamefont {Titum}}, \bibinfo {author} {\bibfnamefont {G.}~\bibnamefont {Pagano}}, \bibinfo {author} {\bibfnamefont {J.}~\bibnamefont {Zhang}}, \bibinfo {author} {\bibfnamefont {C.}~\bibnamefont {Monroe}},\ and\ \bibinfo {author} {\bibfnamefont {A.~V.}\ \bibnamefont {Gorshkov}},\ }\bibfield  {title} {\bibinfo {title} {Confined quasiparticle dynamics in long-range interacting quantum spin chains},\ }\href {https://doi.org/10.1103/PhysRevLett.122.150601} {\bibfield  {journal} {\bibinfo  {journal} {Phys. Rev. Lett.}\ }\textbf {\bibinfo {volume} {122}},\ \bibinfo {pages} {150601} (\bibinfo {year} {2019})}\BibitemShut {NoStop}%
\bibitem [{\citenamefont {Tan}\ \emph {et~al.}(2021)\citenamefont {Tan}, \citenamefont {Becker}, \citenamefont {Liu}, \citenamefont {Pagano}, \citenamefont {Collins}, \citenamefont {De}, \citenamefont {Feng}, \citenamefont {Kaplan}, \citenamefont {Kyprianidis}, \citenamefont {Lundgren}, \citenamefont {Morong}, \citenamefont {Whitsitt}, \citenamefont {Gorshkov},\ and\ \citenamefont {Monroe}}]{LRIM_CONF_expt}%
  \BibitemOpen
  \bibfield  {author} {\bibinfo {author} {\bibfnamefont {W.~L.}\ \bibnamefont {Tan}}, \bibinfo {author} {\bibfnamefont {P.}~\bibnamefont {Becker}}, \bibinfo {author} {\bibfnamefont {F.}~\bibnamefont {Liu}}, \bibinfo {author} {\bibfnamefont {G.}~\bibnamefont {Pagano}}, \bibinfo {author} {\bibfnamefont {K.~S.}\ \bibnamefont {Collins}}, \bibinfo {author} {\bibfnamefont {A.}~\bibnamefont {De}}, \bibinfo {author} {\bibfnamefont {L.}~\bibnamefont {Feng}}, \bibinfo {author} {\bibfnamefont {H.~B.}\ \bibnamefont {Kaplan}}, \bibinfo {author} {\bibfnamefont {A.}~\bibnamefont {Kyprianidis}}, \bibinfo {author} {\bibfnamefont {R.}~\bibnamefont {Lundgren}}, \bibinfo {author} {\bibfnamefont {W.}~\bibnamefont {Morong}}, \bibinfo {author} {\bibfnamefont {S.}~\bibnamefont {Whitsitt}}, \bibinfo {author} {\bibfnamefont {A.~V.}\ \bibnamefont {Gorshkov}},\ and\ \bibinfo {author} {\bibfnamefont {C.}~\bibnamefont {Monroe}},\ }\bibfield  {title} {\bibinfo {title} {Domain-wall confinement and dynamics in a quantum simulator},\ }\href
  {https://doi.org/10.1038/s41567-021-01194-3} {\bibfield  {journal} {\bibinfo  {journal} {Nature Physics}\ }\textbf {\bibinfo {volume} {17}},\ \bibinfo {pages} {742747} (\bibinfo {year} {2021})}\BibitemShut {NoStop}%
\bibitem [{\citenamefont {Kormos}\ \emph {et~al.}(2017)\citenamefont {Kormos}, \citenamefont {Collura}, \citenamefont {Takács},\ and\ \citenamefont {Calabrese}}]{SR_conf}%
  \BibitemOpen
  \bibfield  {author} {\bibinfo {author} {\bibfnamefont {M.}~\bibnamefont {Kormos}}, \bibinfo {author} {\bibfnamefont {M.}~\bibnamefont {Collura}}, \bibinfo {author} {\bibfnamefont {G.}~\bibnamefont {Takács}},\ and\ \bibinfo {author} {\bibfnamefont {P.}~\bibnamefont {Calabrese}},\ }\bibfield  {title} {\bibinfo {title} {Real-time confinement following a quantum quench to a non-integrable model},\ }\href {https://doi.org/10.1038/nphys3934} {\bibfield  {journal} {\bibinfo  {journal} {Nature Physics}\ }\textbf {\bibinfo {volume} {13}},\ \bibinfo {pages} {249246} (\bibinfo {year} {2017})}\BibitemShut {NoStop}%
\bibitem [{\citenamefont {Vovrosh}\ and\ \citenamefont {Knolle}(2021)}]{SR_CONF_expt}%
  \BibitemOpen
  \bibfield  {author} {\bibinfo {author} {\bibfnamefont {J.}~\bibnamefont {Vovrosh}}\ and\ \bibinfo {author} {\bibfnamefont {J.}~\bibnamefont {Knolle}},\ }\bibfield  {title} {\bibinfo {title} {Confinement and entanglement dynamics on a digital quantum computer},\ }\href {https://doi.org/https://doi.org/10.1038/s41598-021-90849-5} {\bibfield  {journal} {\bibinfo  {journal} {Sci Rep}\ }\textbf {\bibinfo {volume} {11}},\ \bibinfo {pages} {11577} (\bibinfo {year} {2021})}\BibitemShut {NoStop}%
\bibitem [{\citenamefont {Lagnese}\ \emph {et~al.}(2021)\citenamefont {Lagnese}, \citenamefont {Surace}, \citenamefont {Kormos},\ and\ \citenamefont {Calabrese}}]{PhysRevB.104.L201106}%
  \BibitemOpen
  \bibfield  {author} {\bibinfo {author} {\bibfnamefont {G.}~\bibnamefont {Lagnese}}, \bibinfo {author} {\bibfnamefont {F.~M.}\ \bibnamefont {Surace}}, \bibinfo {author} {\bibfnamefont {M.}~\bibnamefont {Kormos}},\ and\ \bibinfo {author} {\bibfnamefont {P.}~\bibnamefont {Calabrese}},\ }\bibfield  {title} {\bibinfo {title} {False vacuum decay in quantum spin chains},\ }\href {https://doi.org/10.1103/PhysRevB.104.L201106} {\bibfield  {journal} {\bibinfo  {journal} {Phys. Rev. B}\ }\textbf {\bibinfo {volume} {104}},\ \bibinfo {pages} {L201106} (\bibinfo {year} {2021})}\BibitemShut {NoStop}%
\bibitem [{\citenamefont {Lagnese}\ \emph {et~al.}(2022)\citenamefont {Lagnese}, \citenamefont {Surace}, \citenamefont {Kormos},\ and\ \citenamefont {Calabrese}}]{lagnese2022quenches}%
  \BibitemOpen
  \bibfield  {author} {\bibinfo {author} {\bibfnamefont {G.}~\bibnamefont {Lagnese}}, \bibinfo {author} {\bibfnamefont {F.~M.}\ \bibnamefont {Surace}}, \bibinfo {author} {\bibfnamefont {M.}~\bibnamefont {Kormos}},\ and\ \bibinfo {author} {\bibfnamefont {P.}~\bibnamefont {Calabrese}},\ }\bibfield  {title} {\bibinfo {title} {Quenches and confinement in a heisenberg--ising spin ladder},\ }\href@noop {} {\bibfield  {journal} {\bibinfo  {journal} {Journal of Physics A: Mathematical and Theoretical}\ }\textbf {\bibinfo {volume} {55}},\ \bibinfo {pages} {124003} (\bibinfo {year} {2022})}\BibitemShut {NoStop}%
\bibitem [{\citenamefont {Ranabhat}\ and\ \citenamefont {Collura}(2022)}]{ourown}%
  \BibitemOpen
  \bibfield  {author} {\bibinfo {author} {\bibfnamefont {N.}~\bibnamefont {Ranabhat}}\ and\ \bibinfo {author} {\bibfnamefont {M.}~\bibnamefont {Collura}},\ }\bibfield  {title} {\bibinfo {title} {{Dynamics of the order parameter statistics in the long range Ising model}},\ }\href {https://doi.org/10.21468/SciPostPhys.12.4.126} {\bibfield  {journal} {\bibinfo  {journal} {SciPost Phys.}\ }\textbf {\bibinfo {volume} {12}},\ \bibinfo {pages} {126} (\bibinfo {year} {2022})}\BibitemShut {NoStop}%
\bibitem [{\citenamefont {Ranabhat}\ and\ \citenamefont {Collura}(2024)}]{ranabhat2023thermalization}%
  \BibitemOpen
  \bibfield  {author} {\bibinfo {author} {\bibfnamefont {N.}~\bibnamefont {Ranabhat}}\ and\ \bibinfo {author} {\bibfnamefont {M.}~\bibnamefont {Collura}},\ }\bibfield  {title} {\bibinfo {title} {{Thermalization of long range Ising model in different dynamical regimes: A full counting statistics approach}},\ }\href {https://doi.org/10.21468/SciPostPhysCore.7.2.017} {\bibfield  {journal} {\bibinfo  {journal} {SciPost Phys. Core}\ }\textbf {\bibinfo {volume} {7}},\ \bibinfo {pages} {017} (\bibinfo {year} {2024})}\BibitemShut {NoStop}%
\bibitem [{\citenamefont {Scopa}\ \emph {et~al.}(2022)\citenamefont {Scopa}, \citenamefont {Calabrese},\ and\ \citenamefont {Bastianello}}]{Alvise_conf_entang}%
  \BibitemOpen
  \bibfield  {author} {\bibinfo {author} {\bibfnamefont {S.}~\bibnamefont {Scopa}}, \bibinfo {author} {\bibfnamefont {P.}~\bibnamefont {Calabrese}},\ and\ \bibinfo {author} {\bibfnamefont {A.}~\bibnamefont {Bastianello}},\ }\bibfield  {title} {\bibinfo {title} {Entanglement dynamics in confining spin chains},\ }\href {https://doi.org/10.1103/PhysRevB.105.125413} {\bibfield  {journal} {\bibinfo  {journal} {Phys. Rev. B}\ }\textbf {\bibinfo {volume} {105}},\ \bibinfo {pages} {125413} (\bibinfo {year} {2022})}\BibitemShut {NoStop}%
\bibitem [{\citenamefont {Neyenhuis}\ \emph {et~al.}(2017)\citenamefont {Neyenhuis}, \citenamefont {Zhang}, \citenamefont {Hess}, \citenamefont {Smith}, \citenamefont {Lee}, \citenamefont {Richerme}, \citenamefont {Gong}, \citenamefont {Gorshkov},\ and\ \citenamefont {Monroe}}]{PRETHERMAL_1}%
  \BibitemOpen
  \bibfield  {author} {\bibinfo {author} {\bibfnamefont {B.}~\bibnamefont {Neyenhuis}}, \bibinfo {author} {\bibfnamefont {J.}~\bibnamefont {Zhang}}, \bibinfo {author} {\bibfnamefont {P.~W.}\ \bibnamefont {Hess}}, \bibinfo {author} {\bibfnamefont {J.}~\bibnamefont {Smith}}, \bibinfo {author} {\bibfnamefont {A.~C.}\ \bibnamefont {Lee}}, \bibinfo {author} {\bibfnamefont {P.}~\bibnamefont {Richerme}}, \bibinfo {author} {\bibfnamefont {Z.-X.}\ \bibnamefont {Gong}}, \bibinfo {author} {\bibfnamefont {A.~V.}\ \bibnamefont {Gorshkov}},\ and\ \bibinfo {author} {\bibfnamefont {C.}~\bibnamefont {Monroe}},\ }\bibfield  {title} {\bibinfo {title} {Observation of prethermalization in long-range interacting spin chains},\ }\bibfield  {journal} {\bibinfo  {journal} {Science Advances}\ }\textbf {\bibinfo {volume} {3}},\ \href {https://doi.org/10.1126/sciadv.1700672} {10.1126/sciadv.1700672} (\bibinfo {year} {2017})\BibitemShut {NoStop}%
\bibitem [{\citenamefont {Belavin}\ \emph {et~al.}(1984)\citenamefont {Belavin}, \citenamefont {Polyakov},\ and\ \citenamefont {Zamolodchikov}}]{belavin1984infinite}%
  \BibitemOpen
  \bibfield  {author} {\bibinfo {author} {\bibfnamefont {A.~A.}\ \bibnamefont {Belavin}}, \bibinfo {author} {\bibfnamefont {A.~M.}\ \bibnamefont {Polyakov}},\ and\ \bibinfo {author} {\bibfnamefont {A.~B.}\ \bibnamefont {Zamolodchikov}},\ }\bibfield  {title} {\bibinfo {title} {Infinite conformal symmetry of critical fluctuations in two dimensions},\ }\href@noop {} {\bibfield  {journal} {\bibinfo  {journal} {Journal of Statistical Physics}\ }\textbf {\bibinfo {volume} {34}},\ \bibinfo {pages} {763} (\bibinfo {year} {1984})}\BibitemShut {NoStop}%
\bibitem [{\citenamefont {Henkel}\ and\ \citenamefont {Saleur}(1989)}]{henkel1989two}%
  \BibitemOpen
  \bibfield  {author} {\bibinfo {author} {\bibfnamefont {M.}~\bibnamefont {Henkel}}\ and\ \bibinfo {author} {\bibfnamefont {H.}~\bibnamefont {Saleur}},\ }\bibfield  {title} {\bibinfo {title} {The two-dimensional ising model in the magnetic field: a numerical check of zamolodchikov's conjecture},\ }\href@noop {} {\bibfield  {journal} {\bibinfo  {journal} {Journal of Physics A: Mathematical and General}\ }\textbf {\bibinfo {volume} {22}},\ \bibinfo {pages} {L513} (\bibinfo {year} {1989})}\BibitemShut {NoStop}%
\bibitem [{\citenamefont {Fonseca}\ and\ \citenamefont {Zamolodchikov}(2006)}]{fonseca2006ising}%
  \BibitemOpen
  \bibfield  {author} {\bibinfo {author} {\bibfnamefont {P.}~\bibnamefont {Fonseca}}\ and\ \bibinfo {author} {\bibfnamefont {A.}~\bibnamefont {Zamolodchikov}},\ }\href {https://arxiv.org/abs/hep-th/0612304} {\bibinfo {title} {Ising spectroscopy i: Mesons at t < tc}} (\bibinfo {year} {2006}),\ \Eprint {https://arxiv.org/abs/hep-th/0612304} {arXiv:hep-th/0612304 [hep-th]} \BibitemShut {NoStop}%
\bibitem [{\citenamefont {Ba\~nuls}\ \emph {et~al.}(2023{\natexlab{a}})\citenamefont {Ba\~nuls}, \citenamefont {Heller}, \citenamefont {Jansen}, \citenamefont {Knaute},\ and\ \citenamefont {Svensson}}]{meson_melting}%
  \BibitemOpen
  \bibfield  {author} {\bibinfo {author} {\bibfnamefont {M.~C.}\ \bibnamefont {Ba\~nuls}}, \bibinfo {author} {\bibfnamefont {M.~P.}\ \bibnamefont {Heller}}, \bibinfo {author} {\bibfnamefont {K.}~\bibnamefont {Jansen}}, \bibinfo {author} {\bibfnamefont {J.}~\bibnamefont {Knaute}},\ and\ \bibinfo {author} {\bibfnamefont {V.}~\bibnamefont {Svensson}},\ }\bibfield  {title} {\bibinfo {title} {Quantum information perspective on meson melting},\ }\href {https://doi.org/10.1103/PhysRevD.108.076016} {\bibfield  {journal} {\bibinfo  {journal} {Phys. Rev. D}\ }\textbf {\bibinfo {volume} {108}},\ \bibinfo {pages} {076016} (\bibinfo {year} {2023}{\natexlab{a}})}\BibitemShut {NoStop}%
\bibitem [{\citenamefont {Kebri\ifmmode~\check{c}\else \v{c}\fi{}}\ \emph {et~al.}(2024)\citenamefont {Kebri\ifmmode~\check{c}\else \v{c}\fi{}}, \citenamefont {Halimeh}, \citenamefont {Schollw\"ock},\ and\ \citenamefont {Grusdt}}]{Jadconfinement}%
  \BibitemOpen
  \bibfield  {author} {\bibinfo {author} {\bibfnamefont {M.~c.~v.}\ \bibnamefont {Kebri\ifmmode~\check{c}\else \v{c}\fi{}}}, \bibinfo {author} {\bibfnamefont {J.~C.}\ \bibnamefont {Halimeh}}, \bibinfo {author} {\bibfnamefont {U.}~\bibnamefont {Schollw\"ock}},\ and\ \bibinfo {author} {\bibfnamefont {F.}~\bibnamefont {Grusdt}},\ }\bibfield  {title} {\bibinfo {title} {Confinement in $(1+1)$-dimensional ${\mathbb{z}}_{2}$ lattice gauge theories at finite temperature},\ }\href {https://doi.org/10.1103/PhysRevB.109.245110} {\bibfield  {journal} {\bibinfo  {journal} {Phys. Rev. B}\ }\textbf {\bibinfo {volume} {109}},\ \bibinfo {pages} {245110} (\bibinfo {year} {2024})}\BibitemShut {NoStop}%
\bibitem [{\citenamefont {Buyens}\ \emph {et~al.}(2016)\citenamefont {Buyens}, \citenamefont {Haegeman}, \citenamefont {Verschelde}, \citenamefont {Verstraete},\ and\ \citenamefont {Van~Acoleyen}}]{PhysRevX.6.041040}%
  \BibitemOpen
  \bibfield  {author} {\bibinfo {author} {\bibfnamefont {B.}~\bibnamefont {Buyens}}, \bibinfo {author} {\bibfnamefont {J.}~\bibnamefont {Haegeman}}, \bibinfo {author} {\bibfnamefont {H.}~\bibnamefont {Verschelde}}, \bibinfo {author} {\bibfnamefont {F.}~\bibnamefont {Verstraete}},\ and\ \bibinfo {author} {\bibfnamefont {K.}~\bibnamefont {Van~Acoleyen}},\ }\bibfield  {title} {\bibinfo {title} {Confinement and string breaking for ${\mathrm{qed}}_{2}$ in the hamiltonian picture},\ }\href {https://doi.org/10.1103/PhysRevX.6.041040} {\bibfield  {journal} {\bibinfo  {journal} {Phys. Rev. X}\ }\textbf {\bibinfo {volume} {6}},\ \bibinfo {pages} {041040} (\bibinfo {year} {2016})}\BibitemShut {NoStop}%
\bibitem [{\citenamefont {Aidelsburger}\ \emph {et~al.}(2022)\citenamefont {Aidelsburger}, \citenamefont {Barbiero}, \citenamefont {Bermudez}, \citenamefont {Chanda}, \citenamefont {Dauphin}, \citenamefont {Gonz{\'a}lez-Cuadra}, \citenamefont {Grzybowski}, \citenamefont {Hands}, \citenamefont {Jendrzejewski}, \citenamefont {J{\"u}nemann} \emph {et~al.}}]{aidelsburger2022cold}%
  \BibitemOpen
  \bibfield  {author} {\bibinfo {author} {\bibfnamefont {M.}~\bibnamefont {Aidelsburger}}, \bibinfo {author} {\bibfnamefont {L.}~\bibnamefont {Barbiero}}, \bibinfo {author} {\bibfnamefont {A.}~\bibnamefont {Bermudez}}, \bibinfo {author} {\bibfnamefont {T.}~\bibnamefont {Chanda}}, \bibinfo {author} {\bibfnamefont {A.}~\bibnamefont {Dauphin}}, \bibinfo {author} {\bibfnamefont {D.}~\bibnamefont {Gonz{\'a}lez-Cuadra}}, \bibinfo {author} {\bibfnamefont {P.~R.}\ \bibnamefont {Grzybowski}}, \bibinfo {author} {\bibfnamefont {S.}~\bibnamefont {Hands}}, \bibinfo {author} {\bibfnamefont {F.}~\bibnamefont {Jendrzejewski}}, \bibinfo {author} {\bibfnamefont {J.}~\bibnamefont {J{\"u}nemann}}, \emph {et~al.},\ }\bibfield  {title} {\bibinfo {title} {Cold atoms meet lattice gauge theory},\ }\href@noop {} {\bibfield  {journal} {\bibinfo  {journal} {Philosophical Transactions of the Royal Society A}\ }\textbf {\bibinfo {volume} {380}},\ \bibinfo {pages} {20210064} (\bibinfo {year} {2022})}\BibitemShut {NoStop}%
\bibitem [{\citenamefont {Bǎzǎvan}\ \emph {et~al.}(2024)\citenamefont {Bǎzǎvan}, \citenamefont {Saner}, \citenamefont {Tirrito}, \citenamefont {Araneda}, \citenamefont {Srinivas},\ and\ \citenamefont {Bermudez}}]{buazuavan2023synthetic}%
  \BibitemOpen
  \bibfield  {author} {\bibinfo {author} {\bibfnamefont {O.}~\bibnamefont {Bǎzǎvan}}, \bibinfo {author} {\bibfnamefont {S.}~\bibnamefont {Saner}}, \bibinfo {author} {\bibfnamefont {E.}~\bibnamefont {Tirrito}}, \bibinfo {author} {\bibfnamefont {G.}~\bibnamefont {Araneda}}, \bibinfo {author} {\bibfnamefont {R.}~\bibnamefont {Srinivas}},\ and\ \bibinfo {author} {\bibfnamefont {A.}~\bibnamefont {Bermudez}},\ }\bibfield  {title} {\bibinfo {title} {Synthetic $\mathbb{Z}_{2}$ gauge theories based on parametric excitations of trapped ions},\ }\bibfield  {journal} {\bibinfo  {journal} {Communications Physics}\ }\textbf {\bibinfo {volume} {7}},\ \href {https://doi.org/10.1038/s42005-024-01691-w} {10.1038/s42005-024-01691-w} (\bibinfo {year} {2024})\BibitemShut {NoStop}%
\bibitem [{\citenamefont {Mildenberger}\ \emph {et~al.}(2022)\citenamefont {Mildenberger}, \citenamefont {Mruczkiewicz}, \citenamefont {Halimeh}, \citenamefont {Jiang},\ and\ \citenamefont {Hauke}}]{mildenberger2022probing}%
  \BibitemOpen
  \bibfield  {author} {\bibinfo {author} {\bibfnamefont {J.}~\bibnamefont {Mildenberger}}, \bibinfo {author} {\bibfnamefont {W.}~\bibnamefont {Mruczkiewicz}}, \bibinfo {author} {\bibfnamefont {J.~C.}\ \bibnamefont {Halimeh}}, \bibinfo {author} {\bibfnamefont {Z.}~\bibnamefont {Jiang}},\ and\ \bibinfo {author} {\bibfnamefont {P.}~\bibnamefont {Hauke}},\ }\href {https://arxiv.org/abs/2203.08905} {\bibinfo {title} {Probing confinement in a $\mathbb{Z}_2$ lattice gauge theory on a quantum computer}} (\bibinfo {year} {2022}),\ \Eprint {https://arxiv.org/abs/2203.08905} {arXiv:2203.08905 [quant-ph]} \BibitemShut {NoStop}%
\bibitem [{\citenamefont {Das}\ \emph {et~al.}(2023)\citenamefont {Das}, \citenamefont {Borla},\ and\ \citenamefont {Moroz}}]{LGT_deconf_1}%
  \BibitemOpen
  \bibfield  {author} {\bibinfo {author} {\bibfnamefont {A.}~\bibnamefont {Das}}, \bibinfo {author} {\bibfnamefont {U.}~\bibnamefont {Borla}},\ and\ \bibinfo {author} {\bibfnamefont {S.}~\bibnamefont {Moroz}},\ }\bibfield  {title} {\bibinfo {title} {Fractionalized holes in one-dimensional ${\mathbb{z}}_{2}$ gauge theory coupled to fermion matter: Deconfined dynamics and emergent integrability},\ }\href {https://doi.org/10.1103/PhysRevB.107.064302} {\bibfield  {journal} {\bibinfo  {journal} {Phys. Rev. B}\ }\textbf {\bibinfo {volume} {107}},\ \bibinfo {pages} {064302} (\bibinfo {year} {2023})}\BibitemShut {NoStop}%
\bibitem [{\citenamefont {Gonz\'alez-Cuadra}\ \emph {et~al.}(2020)\citenamefont {Gonz\'alez-Cuadra}, \citenamefont {Tagliacozzo}, \citenamefont {Lewenstein},\ and\ \citenamefont {Bermudez}}]{LGT_deconf_2}%
  \BibitemOpen
  \bibfield  {author} {\bibinfo {author} {\bibfnamefont {D.}~\bibnamefont {Gonz\'alez-Cuadra}}, \bibinfo {author} {\bibfnamefont {L.}~\bibnamefont {Tagliacozzo}}, \bibinfo {author} {\bibfnamefont {M.}~\bibnamefont {Lewenstein}},\ and\ \bibinfo {author} {\bibfnamefont {A.}~\bibnamefont {Bermudez}},\ }\bibfield  {title} {\bibinfo {title} {Robust topological order in fermionic ${\mathbb{z}}_{2}$ gauge theories: From aharonov-bohm instability to soliton-induced deconfinement},\ }\href {https://doi.org/10.1103/PhysRevX.10.041007} {\bibfield  {journal} {\bibinfo  {journal} {Phys. Rev. X}\ }\textbf {\bibinfo {volume} {10}},\ \bibinfo {pages} {041007} (\bibinfo {year} {2020})}\BibitemShut {NoStop}%
\bibitem [{\citenamefont {Lerose}\ \emph {et~al.}(2020)\citenamefont {Lerose}, \citenamefont {Surace}, \citenamefont {Mazza}, \citenamefont {Perfetto}, \citenamefont {Collura},\ and\ \citenamefont {Gambassi}}]{SR_deconf_Alessio}%
  \BibitemOpen
  \bibfield  {author} {\bibinfo {author} {\bibfnamefont {A.}~\bibnamefont {Lerose}}, \bibinfo {author} {\bibfnamefont {F.~M.}\ \bibnamefont {Surace}}, \bibinfo {author} {\bibfnamefont {P.~P.}\ \bibnamefont {Mazza}}, \bibinfo {author} {\bibfnamefont {G.}~\bibnamefont {Perfetto}}, \bibinfo {author} {\bibfnamefont {M.}~\bibnamefont {Collura}},\ and\ \bibinfo {author} {\bibfnamefont {A.}~\bibnamefont {Gambassi}},\ }\bibfield  {title} {\bibinfo {title} {Quasilocalized dynamics from confinement of quantum excitations},\ }\href {https://doi.org/10.1103/PhysRevB.102.041118} {\bibfield  {journal} {\bibinfo  {journal} {Phys. Rev. B}\ }\textbf {\bibinfo {volume} {102}},\ \bibinfo {pages} {041118} (\bibinfo {year} {2020})}\BibitemShut {NoStop}%
\bibitem [{\citenamefont {Ba\~nuls}\ \emph {et~al.}(2023{\natexlab{b}})\citenamefont {Ba\~nuls}, \citenamefont {Heller}, \citenamefont {Jansen}, \citenamefont {Knaute},\ and\ \citenamefont {Svensson}}]{mcbanuls}%
  \BibitemOpen
  \bibfield  {author} {\bibinfo {author} {\bibfnamefont {M.~C.}\ \bibnamefont {Ba\~nuls}}, \bibinfo {author} {\bibfnamefont {M.~P.}\ \bibnamefont {Heller}}, \bibinfo {author} {\bibfnamefont {K.}~\bibnamefont {Jansen}}, \bibinfo {author} {\bibfnamefont {J.}~\bibnamefont {Knaute}},\ and\ \bibinfo {author} {\bibfnamefont {V.}~\bibnamefont {Svensson}},\ }\bibfield  {title} {\bibinfo {title} {Quantum information perspective on meson melting},\ }\href {https://doi.org/10.1103/PhysRevD.108.076016} {\bibfield  {journal} {\bibinfo  {journal} {Phys. Rev. D}\ }\textbf {\bibinfo {volume} {108}},\ \bibinfo {pages} {076016} (\bibinfo {year} {2023}{\natexlab{b}})}\BibitemShut {NoStop}%
\bibitem [{\citenamefont {Birnkammer}\ \emph {et~al.}(2022)\citenamefont {Birnkammer}, \citenamefont {Bastianello},\ and\ \citenamefont {Knap}}]{birnkammer2022prethermalization}%
  \BibitemOpen
  \bibfield  {author} {\bibinfo {author} {\bibfnamefont {S.}~\bibnamefont {Birnkammer}}, \bibinfo {author} {\bibfnamefont {A.}~\bibnamefont {Bastianello}},\ and\ \bibinfo {author} {\bibfnamefont {M.}~\bibnamefont {Knap}},\ }\bibfield  {title} {\bibinfo {title} {Prethermalization in one-dimensional quantum many-body systems with confinement},\ }\href@noop {} {\bibfield  {journal} {\bibinfo  {journal} {Nature Communications}\ }\textbf {\bibinfo {volume} {13}},\ \bibinfo {pages} {7663} (\bibinfo {year} {2022})}\BibitemShut {NoStop}%
\bibitem [{\citenamefont {Defenu}\ \emph {et~al.}(2023)\citenamefont {Defenu}, \citenamefont {Donner}, \citenamefont {Macr\`{\i}}, \citenamefont {Pagano}, \citenamefont {Ruffo},\ and\ \citenamefont {Trombettoni}}]{Defenu2023Review_1}%
  \BibitemOpen
  \bibfield  {author} {\bibinfo {author} {\bibfnamefont {N.}~\bibnamefont {Defenu}}, \bibinfo {author} {\bibfnamefont {T.}~\bibnamefont {Donner}}, \bibinfo {author} {\bibfnamefont {T.}~\bibnamefont {Macr\`{\i}}}, \bibinfo {author} {\bibfnamefont {G.}~\bibnamefont {Pagano}}, \bibinfo {author} {\bibfnamefont {S.}~\bibnamefont {Ruffo}},\ and\ \bibinfo {author} {\bibfnamefont {A.}~\bibnamefont {Trombettoni}},\ }\bibfield  {title} {\bibinfo {title} {Long-range interacting quantum systems},\ }\href {https://doi.org/10.1103/RevModPhys.95.035002} {\bibfield  {journal} {\bibinfo  {journal} {Rev. Mod. Phys.}\ }\textbf {\bibinfo {volume} {95}},\ \bibinfo {pages} {035002} (\bibinfo {year} {2023})}\BibitemShut {NoStop}%
\bibitem [{\citenamefont {Defenu}\ \emph {et~al.}(2024)\citenamefont {Defenu}, \citenamefont {Lerose},\ and\ \citenamefont {Pappalardi}}]{Defenu2023Review_2}%
  \BibitemOpen
  \bibfield  {author} {\bibinfo {author} {\bibfnamefont {N.}~\bibnamefont {Defenu}}, \bibinfo {author} {\bibfnamefont {A.}~\bibnamefont {Lerose}},\ and\ \bibinfo {author} {\bibfnamefont {S.}~\bibnamefont {Pappalardi}},\ }\bibfield  {title} {\bibinfo {title} {Out-of-equilibrium dynamics of quantum many-body systems with long-range interactions},\ }\href {https://doi.org/10.1016/j.physrep.2024.04.005} {\bibfield  {journal} {\bibinfo  {journal} {Physics Reports}\ }\textbf {\bibinfo {volume} {1074}},\ \bibinfo {pages} {1–92} (\bibinfo {year} {2024})}\BibitemShut {NoStop}%
\bibitem [{\citenamefont {Schollw{\"o}ck}(2011)}]{schollwock2011density}%
  \BibitemOpen
  \bibfield  {author} {\bibinfo {author} {\bibfnamefont {U.}~\bibnamefont {Schollw{\"o}ck}},\ }\bibfield  {title} {\bibinfo {title} {The density-matrix renormalization group in the age of matrix product states},\ }\href@noop {} {\bibfield  {journal} {\bibinfo  {journal} {Annals of physics}\ }\textbf {\bibinfo {volume} {326}},\ \bibinfo {pages} {96} (\bibinfo {year} {2011})}\BibitemShut {NoStop}%
\bibitem [{\citenamefont {Or{\'u}s}(2014)}]{orus2014practical}%
  \BibitemOpen
  \bibfield  {author} {\bibinfo {author} {\bibfnamefont {R.}~\bibnamefont {Or{\'u}s}},\ }\bibfield  {title} {\bibinfo {title} {A practical introduction to tensor networks: Matrix product states and projected entangled pair states},\ }\href@noop {} {\bibfield  {journal} {\bibinfo  {journal} {Annals of physics}\ }\textbf {\bibinfo {volume} {349}},\ \bibinfo {pages} {117} (\bibinfo {year} {2014})}\BibitemShut {NoStop}%
\bibitem [{\citenamefont {Ran}\ \emph {et~al.}(2020)\citenamefont {Ran}, \citenamefont {Tirrito}, \citenamefont {Peng}, \citenamefont {Chen}, \citenamefont {Tagliacozzo}, \citenamefont {Su},\ and\ \citenamefont {Lewenstein}}]{ran2020tensor}%
  \BibitemOpen
  \bibfield  {author} {\bibinfo {author} {\bibfnamefont {S.-J.}\ \bibnamefont {Ran}}, \bibinfo {author} {\bibfnamefont {E.}~\bibnamefont {Tirrito}}, \bibinfo {author} {\bibfnamefont {C.}~\bibnamefont {Peng}}, \bibinfo {author} {\bibfnamefont {X.}~\bibnamefont {Chen}}, \bibinfo {author} {\bibfnamefont {L.}~\bibnamefont {Tagliacozzo}}, \bibinfo {author} {\bibfnamefont {G.}~\bibnamefont {Su}},\ and\ \bibinfo {author} {\bibfnamefont {M.}~\bibnamefont {Lewenstein}},\ }\href@noop {} {\emph {\bibinfo {title} {Tensor network contractions: methods and applications to quantum many-body systems}}}\ (\bibinfo  {publisher} {Springer Nature},\ \bibinfo {year} {2020})\BibitemShut {NoStop}%
\bibitem [{\citenamefont {Verstraete}\ \emph {et~al.}(2004)\citenamefont {Verstraete}, \citenamefont {Garc\'{\i}a-Ripoll},\ and\ \citenamefont {Cirac}}]{fintem1}%
  \BibitemOpen
  \bibfield  {author} {\bibinfo {author} {\bibfnamefont {F.}~\bibnamefont {Verstraete}}, \bibinfo {author} {\bibfnamefont {J.~J.}\ \bibnamefont {Garc\'{\i}a-Ripoll}},\ and\ \bibinfo {author} {\bibfnamefont {J.~I.}\ \bibnamefont {Cirac}},\ }\bibfield  {title} {\bibinfo {title} {Matrix product density operators: Simulation of finite-temperature and dissipative systems},\ }\href {https://doi.org/10.1103/PhysRevLett.93.207204} {\bibfield  {journal} {\bibinfo  {journal} {Phys. Rev. Lett.}\ }\textbf {\bibinfo {volume} {93}},\ \bibinfo {pages} {207204} (\bibinfo {year} {2004})}\BibitemShut {NoStop}%
\bibitem [{\citenamefont {Werner}\ \emph {et~al.}(2016)\citenamefont {Werner}, \citenamefont {Jaschke}, \citenamefont {Silvi}, \citenamefont {Kliesch}, \citenamefont {Calarco}, \citenamefont {Eisert},\ and\ \citenamefont {Montangero}}]{fintem2}%
  \BibitemOpen
  \bibfield  {author} {\bibinfo {author} {\bibfnamefont {A.~H.}\ \bibnamefont {Werner}}, \bibinfo {author} {\bibfnamefont {D.}~\bibnamefont {Jaschke}}, \bibinfo {author} {\bibfnamefont {P.}~\bibnamefont {Silvi}}, \bibinfo {author} {\bibfnamefont {M.}~\bibnamefont {Kliesch}}, \bibinfo {author} {\bibfnamefont {T.}~\bibnamefont {Calarco}}, \bibinfo {author} {\bibfnamefont {J.}~\bibnamefont {Eisert}},\ and\ \bibinfo {author} {\bibfnamefont {S.}~\bibnamefont {Montangero}},\ }\bibfield  {title} {\bibinfo {title} {Positive tensor network approach for simulating open quantum many-body systems},\ }\href {https://doi.org/10.1103/PhysRevLett.116.237201} {\bibfield  {journal} {\bibinfo  {journal} {Phys. Rev. Lett.}\ }\textbf {\bibinfo {volume} {116}},\ \bibinfo {pages} {237201} (\bibinfo {year} {2016})}\BibitemShut {NoStop}%
\bibitem [{\citenamefont {Jaschke}\ \emph {et~al.}(2018)\citenamefont {Jaschke}, \citenamefont {Montangero},\ and\ \citenamefont {Carr}}]{LDCarr}%
  \BibitemOpen
  \bibfield  {author} {\bibinfo {author} {\bibfnamefont {D.}~\bibnamefont {Jaschke}}, \bibinfo {author} {\bibfnamefont {S.}~\bibnamefont {Montangero}},\ and\ \bibinfo {author} {\bibfnamefont {L.~D.}\ \bibnamefont {Carr}},\ }\bibfield  {title} {\bibinfo {title} {One-dimensional many-body entangled open quantum systems with tensor network methods},\ }\href {https://doi.org/10.1088/2058-9565/aae724} {\bibfield  {journal} {\bibinfo  {journal} {Quantum Science and Technology}\ }\textbf {\bibinfo {volume} {4}},\ \bibinfo {pages} {013001} (\bibinfo {year} {2018})}\BibitemShut {NoStop}%
\bibitem [{\citenamefont {Zhang}\ \emph {et~al.}(2017{\natexlab{a}})\citenamefont {Zhang}, \citenamefont {Pagano}, \citenamefont {Hess}, \citenamefont {Kyprianidis}, \citenamefont {Becker}, \citenamefont {Kaplan}, \citenamefont {Gorshkov}, \citenamefont {Gong},\ and\ \citenamefont {Monroe}}]{DYN_PT_expt_1}%
  \BibitemOpen
  \bibfield  {author} {\bibinfo {author} {\bibfnamefont {J.}~\bibnamefont {Zhang}}, \bibinfo {author} {\bibfnamefont {G.}~\bibnamefont {Pagano}}, \bibinfo {author} {\bibfnamefont {P.~W.}\ \bibnamefont {Hess}}, \bibinfo {author} {\bibfnamefont {A.}~\bibnamefont {Kyprianidis}}, \bibinfo {author} {\bibfnamefont {P.}~\bibnamefont {Becker}}, \bibinfo {author} {\bibfnamefont {H.}~\bibnamefont {Kaplan}}, \bibinfo {author} {\bibfnamefont {A.~V.}\ \bibnamefont {Gorshkov}}, \bibinfo {author} {\bibfnamefont {Z.-X.}\ \bibnamefont {Gong}},\ and\ \bibinfo {author} {\bibfnamefont {C.}~\bibnamefont {Monroe}},\ }\bibfield  {title} {\bibinfo {title} {Observation of a many-body dynamical phase transition with a 53-qubit quantum simulator},\ }\href {https://doi.org/https://doi.org/10.1038/nature24654} {\bibfield  {journal} {\bibinfo  {journal} {Nature}\ }\textbf {\bibinfo {volume} {551}},\ \bibinfo {pages} {601} (\bibinfo {year} {2017}{\natexlab{a}})}\BibitemShut {NoStop}%
\bibitem [{\citenamefont {Jurcevic}\ \emph {et~al.}(2017)\citenamefont {Jurcevic}, \citenamefont {Shen}, \citenamefont {Hauke}, \citenamefont {Maier}, \citenamefont {Brydges}, \citenamefont {Hempel}, \citenamefont {Lanyon}, \citenamefont {Heyl}, \citenamefont {Blatt},\ and\ \citenamefont {Roos}}]{DYN_PT_expt_2}%
  \BibitemOpen
  \bibfield  {author} {\bibinfo {author} {\bibfnamefont {P.}~\bibnamefont {Jurcevic}}, \bibinfo {author} {\bibfnamefont {H.}~\bibnamefont {Shen}}, \bibinfo {author} {\bibfnamefont {P.}~\bibnamefont {Hauke}}, \bibinfo {author} {\bibfnamefont {C.}~\bibnamefont {Maier}}, \bibinfo {author} {\bibfnamefont {T.}~\bibnamefont {Brydges}}, \bibinfo {author} {\bibfnamefont {C.}~\bibnamefont {Hempel}}, \bibinfo {author} {\bibfnamefont {B.~P.}\ \bibnamefont {Lanyon}}, \bibinfo {author} {\bibfnamefont {M.}~\bibnamefont {Heyl}}, \bibinfo {author} {\bibfnamefont {R.}~\bibnamefont {Blatt}},\ and\ \bibinfo {author} {\bibfnamefont {C.~F.}\ \bibnamefont {Roos}},\ }\bibfield  {title} {\bibinfo {title} {Direct observation of dynamical quantum phase transitions in an interacting many-body system},\ }\href {https://doi.org/10.1103/PhysRevLett.119.080501} {\bibfield  {journal} {\bibinfo  {journal} {Phys. Rev. Lett.}\ }\textbf {\bibinfo {volume} {119}},\ \bibinfo {pages} {080501} (\bibinfo {year} {2017})}\BibitemShut {NoStop}%
\bibitem [{\citenamefont {Blatt}\ and\ \citenamefont {Roos}(2012)}]{LRIM_EXPT1}%
  \BibitemOpen
  \bibfield  {author} {\bibinfo {author} {\bibfnamefont {R.}~\bibnamefont {Blatt}}\ and\ \bibinfo {author} {\bibfnamefont {C.}~\bibnamefont {Roos}},\ }\bibfield  {title} {\bibinfo {title} {Quantum simulations with trapped ions},\ }\href {https://doi.org/https://doi.org/10.1038/nphys2252} {\bibfield  {journal} {\bibinfo  {journal} {Nature}\ }\textbf {\bibinfo {volume} {8}},\ \bibinfo {pages} {277–284} (\bibinfo {year} {2012})}\BibitemShut {NoStop}%
\bibitem [{\citenamefont {Knap}\ \emph {et~al.}(2013{\natexlab{a}})\citenamefont {Knap}, \citenamefont {Kantian}, \citenamefont {Giamarchi}, \citenamefont {Bloch}, \citenamefont {Lukin},\ and\ \citenamefont {Demler}}]{LRIM_EXPT2}%
  \BibitemOpen
  \bibfield  {author} {\bibinfo {author} {\bibfnamefont {M.}~\bibnamefont {Knap}}, \bibinfo {author} {\bibfnamefont {A.}~\bibnamefont {Kantian}}, \bibinfo {author} {\bibfnamefont {T.}~\bibnamefont {Giamarchi}}, \bibinfo {author} {\bibfnamefont {I.}~\bibnamefont {Bloch}}, \bibinfo {author} {\bibfnamefont {M.~D.}\ \bibnamefont {Lukin}},\ and\ \bibinfo {author} {\bibfnamefont {E.}~\bibnamefont {Demler}},\ }\bibfield  {title} {\bibinfo {title} {Probing real-space and time-resolved correlation functions with many-body ramsey interferometry},\ }\href {https://doi.org/10.1103/PhysRevLett.111.147205} {\bibfield  {journal} {\bibinfo  {journal} {Phys. Rev. Lett.}\ }\textbf {\bibinfo {volume} {111}},\ \bibinfo {pages} {147205} (\bibinfo {year} {2013}{\natexlab{a}})}\BibitemShut {NoStop}%
\bibitem [{\citenamefont {Zeiher}\ \emph {et~al.}(2017)\citenamefont {Zeiher}, \citenamefont {Choi}, \citenamefont {Rubio-Abadal}, \citenamefont {Pohl}, \citenamefont {van Bijnen}, \citenamefont {Bloch},\ and\ \citenamefont {Gross}}]{LRIM_EXPT3}%
  \BibitemOpen
  \bibfield  {author} {\bibinfo {author} {\bibfnamefont {J.}~\bibnamefont {Zeiher}}, \bibinfo {author} {\bibfnamefont {J.-y.}\ \bibnamefont {Choi}}, \bibinfo {author} {\bibfnamefont {A.}~\bibnamefont {Rubio-Abadal}}, \bibinfo {author} {\bibfnamefont {T.}~\bibnamefont {Pohl}}, \bibinfo {author} {\bibfnamefont {R.}~\bibnamefont {van Bijnen}}, \bibinfo {author} {\bibfnamefont {I.}~\bibnamefont {Bloch}},\ and\ \bibinfo {author} {\bibfnamefont {C.}~\bibnamefont {Gross}},\ }\bibfield  {title} {\bibinfo {title} {Coherent many-body spin dynamics in a long-range interacting ising chain},\ }\href {https://doi.org/10.1103/PhysRevX.7.041063} {\bibfield  {journal} {\bibinfo  {journal} {Phys. Rev. X}\ }\textbf {\bibinfo {volume} {7}},\ \bibinfo {pages} {041063} (\bibinfo {year} {2017})}\BibitemShut {NoStop}%
\bibitem [{\citenamefont {Bernien}\ \emph {et~al.}(2017)\citenamefont {Bernien}, \citenamefont {Schwartz}, \citenamefont {Keesling},\ and\ \citenamefont {et~al.}}]{LRIM_EXPT4}%
  \BibitemOpen
  \bibfield  {author} {\bibinfo {author} {\bibfnamefont {H.}~\bibnamefont {Bernien}}, \bibinfo {author} {\bibfnamefont {S.}~\bibnamefont {Schwartz}}, \bibinfo {author} {\bibfnamefont {A.}~\bibnamefont {Keesling}},\ and\ \bibinfo {author} {\bibnamefont {et~al.}},\ }\bibfield  {title} {\bibinfo {title} {Probing many-body dynamics on a 51-atom quantum simulator},\ }\href {https://doi.org/https://doi.org/10.1038/nature24622} {\bibfield  {journal} {\bibinfo  {journal} {Nature}\ }\textbf {\bibinfo {volume} {551}},\ \bibinfo {pages} {579–584} (\bibinfo {year} {2017})}\BibitemShut {NoStop}%
\bibitem [{\citenamefont {Richerme}\ \emph {et~al.}(2014)\citenamefont {Richerme}, \citenamefont {Gong}, \citenamefont {Lee},\ and\ \citenamefont {et~al.}}]{LRIM_EXPT5}%
  \BibitemOpen
  \bibfield  {author} {\bibinfo {author} {\bibfnamefont {P.}~\bibnamefont {Richerme}}, \bibinfo {author} {\bibfnamefont {Z.}~\bibnamefont {Gong}}, \bibinfo {author} {\bibfnamefont {A.}~\bibnamefont {Lee}},\ and\ \bibinfo {author} {\bibnamefont {et~al.}},\ }\bibfield  {title} {\bibinfo {title} {Non-local propagation of correlations in quantum systems with long-range interactions},\ }\href {https://doi.org/https://doi.org/10.1038/nature13450} {\bibfield  {journal} {\bibinfo  {journal} {Nature}\ }\textbf {\bibinfo {volume} {511}},\ \bibinfo {pages} {198–201} (\bibinfo {year} {2014})}\BibitemShut {NoStop}%
\bibitem [{\citenamefont {Britton}\ \emph {et~al.}(2012)\citenamefont {Britton}, \citenamefont {Sawyer}, \citenamefont {Keith},\ and\ \citenamefont {et~al.}}]{LRIM_EXPT6}%
  \BibitemOpen
  \bibfield  {author} {\bibinfo {author} {\bibfnamefont {J.}~\bibnamefont {Britton}}, \bibinfo {author} {\bibfnamefont {B.}~\bibnamefont {Sawyer}}, \bibinfo {author} {\bibfnamefont {A.}~\bibnamefont {Keith}},\ and\ \bibinfo {author} {\bibnamefont {et~al.}},\ }\bibfield  {title} {\bibinfo {title} {Engineered two-dimensional ising interactions in a trapped-ion quantum simulator with hundreds of spins},\ }\href {https://doi.org/https://doi.org/10.1038/nature10981} {\bibfield  {journal} {\bibinfo  {journal} {Nature}\ }\textbf {\bibinfo {volume} {484}},\ \bibinfo {pages} {489–492} (\bibinfo {year} {2012})}\BibitemShut {NoStop}%
\bibitem [{\citenamefont {Lieb}\ \emph {et~al.}(1961)\citenamefont {Lieb}, \citenamefont {Schultz},\ and\ \citenamefont {Mattis}}]{TFIM_JW}%
  \BibitemOpen
  \bibfield  {author} {\bibinfo {author} {\bibfnamefont {E.}~\bibnamefont {Lieb}}, \bibinfo {author} {\bibfnamefont {T.}~\bibnamefont {Schultz}},\ and\ \bibinfo {author} {\bibfnamefont {D.}~\bibnamefont {Mattis}},\ }\bibfield  {title} {\bibinfo {title} {Two soluble models of an antiferromagnetic chain},\ }\href {https://doi.org/https://doi.org/10.1016/0003-4916(61)90115-4} {\bibfield  {journal} {\bibinfo  {journal} {Annals of Physics}\ }\textbf {\bibinfo {volume} {16}},\ \bibinfo {pages} {407} (\bibinfo {year} {1961})}\BibitemShut {NoStop}%
\bibitem [{\citenamefont {\ifmmode \check{Z}\else \v{Z}\fi{}unkovi\ifmmode~\check{c}\else \v{c}\fi{}}\ \emph {et~al.}(2016)\citenamefont {\ifmmode \check{Z}\else \v{Z}\fi{}unkovi\ifmmode~\check{c}\else \v{c}\fi{}}, \citenamefont {Silva},\ and\ \citenamefont {Fabrizio}}]{FC1}%
  \BibitemOpen
  \bibfield  {author} {\bibinfo {author} {\bibfnamefont {B.}~\bibnamefont {\ifmmode \check{Z}\else \v{Z}\fi{}unkovi\ifmmode~\check{c}\else \v{c}\fi{}}}, \bibinfo {author} {\bibfnamefont {A.}~\bibnamefont {Silva}},\ and\ \bibinfo {author} {\bibfnamefont {M.}~\bibnamefont {Fabrizio}},\ }\bibfield  {title} {\bibinfo {title} {{Dynamical phase transitions and Loschmidt echo in the infinite-range XY model}},\ }\bibfield  {journal} {\bibinfo  {journal} {Phil. Trans. R. Soc. A.}\ }\textbf {\bibinfo {volume} {374}},\ \href {https://doi.org/10.1098/rsta.2015.0160} {10.1098/rsta.2015.0160} (\bibinfo {year} {2016})\BibitemShut {NoStop}%
\bibitem [{\citenamefont {Das}\ \emph {et~al.}(2006)\citenamefont {Das}, \citenamefont {Sengupta}, \citenamefont {Sen},\ and\ \citenamefont {Chakrabarti}}]{FC2}%
  \BibitemOpen
  \bibfield  {author} {\bibinfo {author} {\bibfnamefont {A.}~\bibnamefont {Das}}, \bibinfo {author} {\bibfnamefont {K.}~\bibnamefont {Sengupta}}, \bibinfo {author} {\bibfnamefont {D.}~\bibnamefont {Sen}},\ and\ \bibinfo {author} {\bibfnamefont {B.~K.}\ \bibnamefont {Chakrabarti}},\ }\bibfield  {title} {\bibinfo {title} {Infinite-range ising ferromagnet in a time-dependent transverse magnetic field: Quench and ac dynamics near the quantum critical point},\ }\href {https://doi.org/10.1103/PhysRevB.74.144423} {\bibfield  {journal} {\bibinfo  {journal} {Phys. Rev. B}\ }\textbf {\bibinfo {volume} {74}},\ \bibinfo {pages} {144423} (\bibinfo {year} {2006})}\BibitemShut {NoStop}%
\bibitem [{\citenamefont {Knap}\ \emph {et~al.}(2013{\natexlab{b}})\citenamefont {Knap}, \citenamefont {Kantian}, \citenamefont {Giamarchi}, \citenamefont {Bloch}, \citenamefont {Lukin},\ and\ \citenamefont {Demler}}]{QPT1}%
  \BibitemOpen
  \bibfield  {author} {\bibinfo {author} {\bibfnamefont {M.}~\bibnamefont {Knap}}, \bibinfo {author} {\bibfnamefont {A.}~\bibnamefont {Kantian}}, \bibinfo {author} {\bibfnamefont {T.}~\bibnamefont {Giamarchi}}, \bibinfo {author} {\bibfnamefont {I.}~\bibnamefont {Bloch}}, \bibinfo {author} {\bibfnamefont {M.~D.}\ \bibnamefont {Lukin}},\ and\ \bibinfo {author} {\bibfnamefont {E.}~\bibnamefont {Demler}},\ }\bibfield  {title} {\bibinfo {title} {Probing real-space and time-resolved correlation functions with many-body ramsey interferometry},\ }\href {https://doi.org/10.1103/PhysRevLett.111.147205} {\bibfield  {journal} {\bibinfo  {journal} {Phys. Rev. Lett.}\ }\textbf {\bibinfo {volume} {111}},\ \bibinfo {pages} {147205} (\bibinfo {year} {2013}{\natexlab{b}})}\BibitemShut {NoStop}%
\bibitem [{\citenamefont {Koffel}\ \emph {et~al.}(2012)\citenamefont {Koffel}, \citenamefont {Lewenstein},\ and\ \citenamefont {Tagliacozzo}}]{QPT2}%
  \BibitemOpen
  \bibfield  {author} {\bibinfo {author} {\bibfnamefont {T.}~\bibnamefont {Koffel}}, \bibinfo {author} {\bibfnamefont {M.}~\bibnamefont {Lewenstein}},\ and\ \bibinfo {author} {\bibfnamefont {L.}~\bibnamefont {Tagliacozzo}},\ }\bibfield  {title} {\bibinfo {title} {Entanglement entropy for the long-range ising chain in a transverse field},\ }\href {https://doi.org/10.1103/PhysRevLett.109.267203} {\bibfield  {journal} {\bibinfo  {journal} {Phys. Rev. Lett.}\ }\textbf {\bibinfo {volume} {109}},\ \bibinfo {pages} {267203} (\bibinfo {year} {2012})}\BibitemShut {NoStop}%
\bibitem [{\citenamefont {Fey}\ and\ \citenamefont {Schmidt}(2016)}]{QPT3}%
  \BibitemOpen
  \bibfield  {author} {\bibinfo {author} {\bibfnamefont {S.}~\bibnamefont {Fey}}\ and\ \bibinfo {author} {\bibfnamefont {K.~P.}\ \bibnamefont {Schmidt}},\ }\bibfield  {title} {\bibinfo {title} {Critical behavior of quantum magnets with long-range interactions in the thermodynamic limit},\ }\href {https://doi.org/10.1103/PhysRevB.94.075156} {\bibfield  {journal} {\bibinfo  {journal} {Phys. Rev. B}\ }\textbf {\bibinfo {volume} {94}},\ \bibinfo {pages} {075156} (\bibinfo {year} {2016})}\BibitemShut {NoStop}%
\bibitem [{\citenamefont {Dutta}\ and\ \citenamefont {Bhattacharjee}(2001)}]{LRIM_T_trans1}%
  \BibitemOpen
  \bibfield  {author} {\bibinfo {author} {\bibfnamefont {A.}~\bibnamefont {Dutta}}\ and\ \bibinfo {author} {\bibfnamefont {J.~K.}\ \bibnamefont {Bhattacharjee}},\ }\bibfield  {title} {\bibinfo {title} {Phase transitions in the quantum ising and rotor models with a long-range interaction},\ }\href {https://doi.org/10.1103/PhysRevB.64.184106} {\bibfield  {journal} {\bibinfo  {journal} {Phys. Rev. B}\ }\textbf {\bibinfo {volume} {64}},\ \bibinfo {pages} {184106} (\bibinfo {year} {2001})}\BibitemShut {NoStop}%
\bibitem [{\citenamefont {Gonzalez-Lazo}\ \emph {et~al.}(2021)\citenamefont {Gonzalez-Lazo}, \citenamefont {Heyl}, \citenamefont {Dalmonte},\ and\ \citenamefont {Angelone}}]{LRIM_T_trans2}%
  \BibitemOpen
  \bibfield  {author} {\bibinfo {author} {\bibfnamefont {E.}~\bibnamefont {Gonzalez-Lazo}}, \bibinfo {author} {\bibfnamefont {M.}~\bibnamefont {Heyl}}, \bibinfo {author} {\bibfnamefont {M.}~\bibnamefont {Dalmonte}},\ and\ \bibinfo {author} {\bibfnamefont {A.}~\bibnamefont {Angelone}},\ }\bibfield  {title} {\bibinfo {title} {{Finite-temperature critical behavior of long-range quantum Ising models}},\ }\href {https://doi.org/10.21468/SciPostPhys.11.4.076} {\bibfield  {journal} {\bibinfo  {journal} {SciPost Phys.}\ }\textbf {\bibinfo {volume} {11}},\ \bibinfo {pages} {76} (\bibinfo {year} {2021})}\BibitemShut {NoStop}%
\bibitem [{\citenamefont {\ifmmode \check{Z}\else \v{Z}\fi{}unkovi\ifmmode~\check{c}\else \v{c}\fi{}}\ \emph {et~al.}(2018)\citenamefont {\ifmmode \check{Z}\else \v{Z}\fi{}unkovi\ifmmode~\check{c}\else \v{c}\fi{}}, \citenamefont {Heyl}, \citenamefont {Knap},\ and\ \citenamefont {Silva}}]{DYN_PT_1}%
  \BibitemOpen
  \bibfield  {author} {\bibinfo {author} {\bibfnamefont {B.}~\bibnamefont {\ifmmode \check{Z}\else \v{Z}\fi{}unkovi\ifmmode~\check{c}\else \v{c}\fi{}}}, \bibinfo {author} {\bibfnamefont {M.}~\bibnamefont {Heyl}}, \bibinfo {author} {\bibfnamefont {M.}~\bibnamefont {Knap}},\ and\ \bibinfo {author} {\bibfnamefont {A.}~\bibnamefont {Silva}},\ }\bibfield  {title} {\bibinfo {title} {Dynamical quantum phase transitions in spin chains with long-range interactions: Merging different concepts of nonequilibrium criticality},\ }\href {https://doi.org/10.1103/PhysRevLett.120.130601} {\bibfield  {journal} {\bibinfo  {journal} {Phys. Rev. Lett.}\ }\textbf {\bibinfo {volume} {120}},\ \bibinfo {pages} {130601} (\bibinfo {year} {2018})}\BibitemShut {NoStop}%
\bibitem [{\citenamefont {Halimeh}\ and\ \citenamefont {Zauner-Stauber}(2017)}]{DYN_PT_2}%
  \BibitemOpen
  \bibfield  {author} {\bibinfo {author} {\bibfnamefont {J.~C.}\ \bibnamefont {Halimeh}}\ and\ \bibinfo {author} {\bibfnamefont {V.}~\bibnamefont {Zauner-Stauber}},\ }\bibfield  {title} {\bibinfo {title} {Dynamical phase diagram of quantum spin chains with long-range interactions},\ }\href {https://doi.org/10.1103/PhysRevB.96.134427} {\bibfield  {journal} {\bibinfo  {journal} {Phys. Rev. B}\ }\textbf {\bibinfo {volume} {96}},\ \bibinfo {pages} {134427} (\bibinfo {year} {2017})}\BibitemShut {NoStop}%
\bibitem [{\citenamefont {Khasseh}\ \emph {et~al.}(2020)\citenamefont {Khasseh}, \citenamefont {Russomanno}, \citenamefont {Schmitt}, \citenamefont {Heyl},\ and\ \citenamefont {Fazio}}]{DYN_PT_3}%
  \BibitemOpen
  \bibfield  {author} {\bibinfo {author} {\bibfnamefont {R.}~\bibnamefont {Khasseh}}, \bibinfo {author} {\bibfnamefont {A.}~\bibnamefont {Russomanno}}, \bibinfo {author} {\bibfnamefont {M.}~\bibnamefont {Schmitt}}, \bibinfo {author} {\bibfnamefont {M.}~\bibnamefont {Heyl}},\ and\ \bibinfo {author} {\bibfnamefont {R.}~\bibnamefont {Fazio}},\ }\bibfield  {title} {\bibinfo {title} {Discrete truncated wigner approach to dynamical phase transitions in ising models after a quantum quench},\ }\href {https://doi.org/10.1103/PhysRevB.102.014303} {\bibfield  {journal} {\bibinfo  {journal} {Phys. Rev. B}\ }\textbf {\bibinfo {volume} {102}},\ \bibinfo {pages} {014303} (\bibinfo {year} {2020})}\BibitemShut {NoStop}%
\bibitem [{\citenamefont {Homrighausen}\ \emph {et~al.}(2017)\citenamefont {Homrighausen}, \citenamefont {Abeling}, \citenamefont {Zauner-Stauber},\ and\ \citenamefont {Halimeh}}]{Halimeh_2}%
  \BibitemOpen
  \bibfield  {author} {\bibinfo {author} {\bibfnamefont {I.}~\bibnamefont {Homrighausen}}, \bibinfo {author} {\bibfnamefont {N.~O.}\ \bibnamefont {Abeling}}, \bibinfo {author} {\bibfnamefont {V.}~\bibnamefont {Zauner-Stauber}},\ and\ \bibinfo {author} {\bibfnamefont {J.~C.}\ \bibnamefont {Halimeh}},\ }\bibfield  {title} {\bibinfo {title} {Anomalous dynamical phase in quantum spin chains with long-range interactions},\ }\href {https://doi.org/10.1103/PhysRevB.96.104436} {\bibfield  {journal} {\bibinfo  {journal} {Phys. Rev. B}\ }\textbf {\bibinfo {volume} {96}},\ \bibinfo {pages} {104436} (\bibinfo {year} {2017})}\BibitemShut {NoStop}%
\bibitem [{\citenamefont {Lang}\ \emph {et~al.}(2018)\citenamefont {Lang}, \citenamefont {Frank},\ and\ \citenamefont {Halimeh}}]{Halimeh_3}%
  \BibitemOpen
  \bibfield  {author} {\bibinfo {author} {\bibfnamefont {J.}~\bibnamefont {Lang}}, \bibinfo {author} {\bibfnamefont {B.}~\bibnamefont {Frank}},\ and\ \bibinfo {author} {\bibfnamefont {J.~C.}\ \bibnamefont {Halimeh}},\ }\bibfield  {title} {\bibinfo {title} {Concurrence of dynamical phase transitions at finite temperature in the fully connected transverse-field ising model},\ }\href {https://doi.org/10.1103/PhysRevB.97.174401} {\bibfield  {journal} {\bibinfo  {journal} {Phys. Rev. B}\ }\textbf {\bibinfo {volume} {97}},\ \bibinfo {pages} {174401} (\bibinfo {year} {2018})}\BibitemShut {NoStop}%
\bibitem [{\citenamefont {Halimeh}\ \emph {et~al.}(2017)\citenamefont {Halimeh}, \citenamefont {Zauner-Stauber}, \citenamefont {McCulloch}, \citenamefont {de~Vega}, \citenamefont {Schollw\"ock},\ and\ \citenamefont {Kastner}}]{Halimeh_1}%
  \BibitemOpen
  \bibfield  {author} {\bibinfo {author} {\bibfnamefont {J.~C.}\ \bibnamefont {Halimeh}}, \bibinfo {author} {\bibfnamefont {V.}~\bibnamefont {Zauner-Stauber}}, \bibinfo {author} {\bibfnamefont {I.~P.}\ \bibnamefont {McCulloch}}, \bibinfo {author} {\bibfnamefont {I.}~\bibnamefont {de~Vega}}, \bibinfo {author} {\bibfnamefont {U.}~\bibnamefont {Schollw\"ock}},\ and\ \bibinfo {author} {\bibfnamefont {M.}~\bibnamefont {Kastner}},\ }\bibfield  {title} {\bibinfo {title} {Prethermalization and persistent order in the absence of a thermal phase transition},\ }\href {https://doi.org/10.1103/PhysRevB.95.024302} {\bibfield  {journal} {\bibinfo  {journal} {Phys. Rev. B}\ }\textbf {\bibinfo {volume} {95}},\ \bibinfo {pages} {024302} (\bibinfo {year} {2017})}\BibitemShut {NoStop}%
\bibitem [{\citenamefont {Foss-Feig}\ \emph {et~al.}(2015)\citenamefont {Foss-Feig}, \citenamefont {Gong}, \citenamefont {Clark},\ and\ \citenamefont {Gorshkov}}]{LIGHT_CONE_1}%
  \BibitemOpen
  \bibfield  {author} {\bibinfo {author} {\bibfnamefont {M.}~\bibnamefont {Foss-Feig}}, \bibinfo {author} {\bibfnamefont {Z.-X.}\ \bibnamefont {Gong}}, \bibinfo {author} {\bibfnamefont {C.~W.}\ \bibnamefont {Clark}},\ and\ \bibinfo {author} {\bibfnamefont {A.~V.}\ \bibnamefont {Gorshkov}},\ }\bibfield  {title} {\bibinfo {title} {Nearly linear light cones in long-range interacting quantum systems},\ }\href {https://doi.org/10.1103/PhysRevLett.114.157201} {\bibfield  {journal} {\bibinfo  {journal} {Phys. Rev. Lett.}\ }\textbf {\bibinfo {volume} {114}},\ \bibinfo {pages} {157201} (\bibinfo {year} {2015})}\BibitemShut {NoStop}%
\bibitem [{\citenamefont {Buyskikh}\ \emph {et~al.}(2016)\citenamefont {Buyskikh}, \citenamefont {Fagotti}, \citenamefont {Schachenmayer}, \citenamefont {Essler},\ and\ \citenamefont {Daley}}]{LIGHT_CONE_2}%
  \BibitemOpen
  \bibfield  {author} {\bibinfo {author} {\bibfnamefont {A.~S.}\ \bibnamefont {Buyskikh}}, \bibinfo {author} {\bibfnamefont {M.}~\bibnamefont {Fagotti}}, \bibinfo {author} {\bibfnamefont {J.}~\bibnamefont {Schachenmayer}}, \bibinfo {author} {\bibfnamefont {F.}~\bibnamefont {Essler}},\ and\ \bibinfo {author} {\bibfnamefont {A.~J.}\ \bibnamefont {Daley}},\ }\bibfield  {title} {\bibinfo {title} {Entanglement growth and correlation spreading with variable-range interactions in spin and fermionic tunneling models},\ }\href {https://doi.org/10.1103/PhysRevA.93.053620} {\bibfield  {journal} {\bibinfo  {journal} {Phys. Rev. A}\ }\textbf {\bibinfo {volume} {93}},\ \bibinfo {pages} {053620} (\bibinfo {year} {2016})}\BibitemShut {NoStop}%
\bibitem [{\citenamefont {Kyprianidis}\ \emph {et~al.}(2021)\citenamefont {Kyprianidis}, \citenamefont {Machado}, \citenamefont {Morong}, \citenamefont {Becker}, \citenamefont {Collins}, \citenamefont {ELSE}, \citenamefont {Feng}, \citenamefont {Hess}, \citenamefont {Nayak}, \citenamefont {Pagano}, \citenamefont {Yao},\ and\ \citenamefont {Monroe}}]{DTC1}%
  \BibitemOpen
  \bibfield  {author} {\bibinfo {author} {\bibfnamefont {A.}~\bibnamefont {Kyprianidis}}, \bibinfo {author} {\bibfnamefont {F.}~\bibnamefont {Machado}}, \bibinfo {author} {\bibfnamefont {W.}~\bibnamefont {Morong}}, \bibinfo {author} {\bibfnamefont {P.}~\bibnamefont {Becker}}, \bibinfo {author} {\bibfnamefont {K.}~\bibnamefont {Collins}}, \bibinfo {author} {\bibfnamefont {D.}~\bibnamefont {ELSE}}, \bibinfo {author} {\bibfnamefont {L.}~\bibnamefont {Feng}}, \bibinfo {author} {\bibfnamefont {P.}~\bibnamefont {Hess}}, \bibinfo {author} {\bibfnamefont {C.}~\bibnamefont {Nayak}}, \bibinfo {author} {\bibfnamefont {G.}~\bibnamefont {Pagano}}, \bibinfo {author} {\bibfnamefont {N.}~\bibnamefont {Yao}},\ and\ \bibinfo {author} {\bibfnamefont {C.}~\bibnamefont {Monroe}},\ }\bibfield  {title} {\bibinfo {title} {Observation of a prethermal discrete time crystal},\ }\href {https://doi.org/10.1126/science.abg8102} {\bibfield  {journal} {\bibinfo  {journal} {Science}\ }\textbf {\bibinfo {volume} {372}},\ \bibinfo {pages}
  {1192} (\bibinfo {year} {2021})}\BibitemShut {NoStop}%
\bibitem [{\citenamefont {Zhang}\ \emph {et~al.}(2017{\natexlab{b}})\citenamefont {Zhang}, \citenamefont {Hess}, \citenamefont {Kyprianidis}, \citenamefont {Becker}, \citenamefont {Lee}, \citenamefont {Smith}, \citenamefont {Pagano}, \citenamefont {Potirniche}, \citenamefont {Porter}, \citenamefont {Vishwanath}, \citenamefont {Yao},\ and\ \citenamefont {Monroe}}]{DTC2}%
  \BibitemOpen
  \bibfield  {author} {\bibinfo {author} {\bibfnamefont {J.}~\bibnamefont {Zhang}}, \bibinfo {author} {\bibfnamefont {P.~W.}\ \bibnamefont {Hess}}, \bibinfo {author} {\bibfnamefont {A.}~\bibnamefont {Kyprianidis}}, \bibinfo {author} {\bibfnamefont {P.}~\bibnamefont {Becker}}, \bibinfo {author} {\bibfnamefont {A.}~\bibnamefont {Lee}}, \bibinfo {author} {\bibfnamefont {J.}~\bibnamefont {Smith}}, \bibinfo {author} {\bibfnamefont {G.}~\bibnamefont {Pagano}}, \bibinfo {author} {\bibfnamefont {I.~D.}\ \bibnamefont {Potirniche}}, \bibinfo {author} {\bibfnamefont {A.~C.}\ \bibnamefont {Porter}}, \bibinfo {author} {\bibfnamefont {A.}~\bibnamefont {Vishwanath}}, \bibinfo {author} {\bibfnamefont {N.~Y.}\ \bibnamefont {Yao}},\ and\ \bibinfo {author} {\bibfnamefont {C.}~\bibnamefont {Monroe}},\ }\bibfield  {title} {\bibinfo {title} {Observation of a discrete time crystal},\ }\href {https://doi.org/https://doi.org/10.1038/nature21413} {\bibfield  {journal} {\bibinfo  {journal} {Nature}\ }\textbf {\bibinfo {volume} {543}},\
  \bibinfo {pages} {217} (\bibinfo {year} {2017}{\natexlab{b}})}\BibitemShut {NoStop}%
\bibitem [{\citenamefont {Haegeman}\ \emph {et~al.}(2011)\citenamefont {Haegeman}, \citenamefont {Cirac}, \citenamefont {Osborne}, \citenamefont {Pi\ifmmode~\check{z}\else \v{z}\fi{}orn}, \citenamefont {Verschelde},\ and\ \citenamefont {Verstraete}}]{TDVP_one}%
  \BibitemOpen
  \bibfield  {author} {\bibinfo {author} {\bibfnamefont {J.}~\bibnamefont {Haegeman}}, \bibinfo {author} {\bibfnamefont {J.~I.}\ \bibnamefont {Cirac}}, \bibinfo {author} {\bibfnamefont {T.~J.}\ \bibnamefont {Osborne}}, \bibinfo {author} {\bibfnamefont {I.}~\bibnamefont {Pi\ifmmode~\check{z}\else \v{z}\fi{}orn}}, \bibinfo {author} {\bibfnamefont {H.}~\bibnamefont {Verschelde}},\ and\ \bibinfo {author} {\bibfnamefont {F.}~\bibnamefont {Verstraete}},\ }\bibfield  {title} {\bibinfo {title} {Time-dependent variational principle for quantum lattices},\ }\href {https://doi.org/10.1103/PhysRevLett.107.070601} {\bibfield  {journal} {\bibinfo  {journal} {Phys. Rev. Lett.}\ }\textbf {\bibinfo {volume} {107}},\ \bibinfo {pages} {070601} (\bibinfo {year} {2011})}\BibitemShut {NoStop}%
\bibitem [{\citenamefont {Haegeman}\ \emph {et~al.}(2016)\citenamefont {Haegeman}, \citenamefont {Lubich}, \citenamefont {Oseledets}, \citenamefont {Vandereycken},\ and\ \citenamefont {Verstraete}}]{TDVP_two}%
  \BibitemOpen
  \bibfield  {author} {\bibinfo {author} {\bibfnamefont {J.}~\bibnamefont {Haegeman}}, \bibinfo {author} {\bibfnamefont {C.}~\bibnamefont {Lubich}}, \bibinfo {author} {\bibfnamefont {I.}~\bibnamefont {Oseledets}}, \bibinfo {author} {\bibfnamefont {B.}~\bibnamefont {Vandereycken}},\ and\ \bibinfo {author} {\bibfnamefont {F.}~\bibnamefont {Verstraete}},\ }\bibfield  {title} {\bibinfo {title} {Unifying time evolution and optimization with matrix product states},\ }\href {https://doi.org/10.1103/PhysRevB.94.165116} {\bibfield  {journal} {\bibinfo  {journal} {Phys. Rev. B}\ }\textbf {\bibinfo {volume} {94}},\ \bibinfo {pages} {165116} (\bibinfo {year} {2016})}\BibitemShut {NoStop}%
\bibitem [{\citenamefont {Essler}\ and\ \citenamefont {Fagotti}(2016)}]{Essler_2016}%
  \BibitemOpen
  \bibfield  {author} {\bibinfo {author} {\bibfnamefont {F.~H.~L.}\ \bibnamefont {Essler}}\ and\ \bibinfo {author} {\bibfnamefont {M.}~\bibnamefont {Fagotti}},\ }\bibfield  {title} {\bibinfo {title} {Quench dynamics and relaxation in isolated integrable quantum spin chains},\ }\href {https://doi.org/10.1088/1742-5468/2016/06/064002} {\bibfield  {journal} {\bibinfo  {journal} {Journal of Statistical Mechanics: Theory and Experiment}\ }\textbf {\bibinfo {volume} {2016}},\ \bibinfo {pages} {064002} (\bibinfo {year} {2016})}\BibitemShut {NoStop}%
\bibitem [{\citenamefont {Ba\~nuls}\ \emph {et~al.}(2011)\citenamefont {Ba\~nuls}, \citenamefont {Cirac},\ and\ \citenamefont {Hastings}}]{SandW_therm}%
  \BibitemOpen
  \bibfield  {author} {\bibinfo {author} {\bibfnamefont {M.~C.}\ \bibnamefont {Ba\~nuls}}, \bibinfo {author} {\bibfnamefont {J.~I.}\ \bibnamefont {Cirac}},\ and\ \bibinfo {author} {\bibfnamefont {M.~B.}\ \bibnamefont {Hastings}},\ }\bibfield  {title} {\bibinfo {title} {Strong and weak thermalization of infinite nonintegrable quantum systems},\ }\href {https://doi.org/10.1103/PhysRevLett.106.050405} {\bibfield  {journal} {\bibinfo  {journal} {Phys. Rev. Lett.}\ }\textbf {\bibinfo {volume} {106}},\ \bibinfo {pages} {050405} (\bibinfo {year} {2011})}\BibitemShut {NoStop}%
\bibitem [{\citenamefont {Mazza}\ \emph {et~al.}(2019)\citenamefont {Mazza}, \citenamefont {Perfetto}, \citenamefont {Lerose}, \citenamefont {Collura},\ and\ \citenamefont {Gambassi}}]{single_kink1}%
  \BibitemOpen
  \bibfield  {author} {\bibinfo {author} {\bibfnamefont {P.~P.}\ \bibnamefont {Mazza}}, \bibinfo {author} {\bibfnamefont {G.}~\bibnamefont {Perfetto}}, \bibinfo {author} {\bibfnamefont {A.}~\bibnamefont {Lerose}}, \bibinfo {author} {\bibfnamefont {M.}~\bibnamefont {Collura}},\ and\ \bibinfo {author} {\bibfnamefont {A.}~\bibnamefont {Gambassi}},\ }\bibfield  {title} {\bibinfo {title} {Suppression of transport in nondisordered quantum spin chains due to confined excitations},\ }\href {https://doi.org/10.1103/PhysRevB.99.180302} {\bibfield  {journal} {\bibinfo  {journal} {Phys. Rev. B}\ }\textbf {\bibinfo {volume} {99}},\ \bibinfo {pages} {180302} (\bibinfo {year} {2019})}\BibitemShut {NoStop}%
\bibitem [{\citenamefont {Eisert}\ \emph {et~al.}(2010)\citenamefont {Eisert}, \citenamefont {Cramer},\ and\ \citenamefont {Plenio}}]{area_law_1}%
  \BibitemOpen
  \bibfield  {author} {\bibinfo {author} {\bibfnamefont {J.}~\bibnamefont {Eisert}}, \bibinfo {author} {\bibfnamefont {M.}~\bibnamefont {Cramer}},\ and\ \bibinfo {author} {\bibfnamefont {M.~B.}\ \bibnamefont {Plenio}},\ }\bibfield  {title} {\bibinfo {title} {Colloquium: Area laws for the entanglement entropy},\ }\href {https://doi.org/10.1103/RevModPhys.82.277} {\bibfield  {journal} {\bibinfo  {journal} {Rev. Mod. Phys.}\ }\textbf {\bibinfo {volume} {82}},\ \bibinfo {pages} {277} (\bibinfo {year} {2010})}\BibitemShut {NoStop}%
\bibitem [{\citenamefont {Srednicki}(1993)}]{area_law_3}%
  \BibitemOpen
  \bibfield  {author} {\bibinfo {author} {\bibfnamefont {M.}~\bibnamefont {Srednicki}},\ }\bibfield  {title} {\bibinfo {title} {Entropy and area},\ }\href {https://doi.org/10.1103/PhysRevLett.71.666} {\bibfield  {journal} {\bibinfo  {journal} {Phys. Rev. Lett.}\ }\textbf {\bibinfo {volume} {71}},\ \bibinfo {pages} {666} (\bibinfo {year} {1993})}\BibitemShut {NoStop}%
\bibitem [{\citenamefont {Plenio}\ \emph {et~al.}(2005)\citenamefont {Plenio}, \citenamefont {Eisert}, \citenamefont {Drei\ss{}ig},\ and\ \citenamefont {Cramer}}]{area_law_5}%
  \BibitemOpen
  \bibfield  {author} {\bibinfo {author} {\bibfnamefont {M.~B.}\ \bibnamefont {Plenio}}, \bibinfo {author} {\bibfnamefont {J.}~\bibnamefont {Eisert}}, \bibinfo {author} {\bibfnamefont {J.}~\bibnamefont {Drei\ss{}ig}},\ and\ \bibinfo {author} {\bibfnamefont {M.}~\bibnamefont {Cramer}},\ }\bibfield  {title} {\bibinfo {title} {Entropy, entanglement, and area: Analytical results for harmonic lattice systems},\ }\href {https://doi.org/10.1103/PhysRevLett.94.060503} {\bibfield  {journal} {\bibinfo  {journal} {Phys. Rev. Lett.}\ }\textbf {\bibinfo {volume} {94}},\ \bibinfo {pages} {060503} (\bibinfo {year} {2005})}\BibitemShut {NoStop}%
\bibitem [{\citenamefont {Calabrese}\ and\ \citenamefont {Cardy}(2004)}]{area_law_6}%
  \BibitemOpen
  \bibfield  {author} {\bibinfo {author} {\bibfnamefont {P.}~\bibnamefont {Calabrese}}\ and\ \bibinfo {author} {\bibfnamefont {J.}~\bibnamefont {Cardy}},\ }\bibfield  {title} {\bibinfo {title} {Entanglement entropy and quantum field theory},\ }\href {https://doi.org/10.1088/1742-5468/2004/06/p06002} {\bibfield  {journal} {\bibinfo  {journal} {Journal of Statistical Mechanics: Theory and Experiment}\ }\textbf {\bibinfo {volume} {2004}},\ \bibinfo {pages} {P06002} (\bibinfo {year} {2004})}\BibitemShut {NoStop}%
\bibitem [{\citenamefont {Vidal}\ \emph {et~al.}(2003)\citenamefont {Vidal}, \citenamefont {Latorre}, \citenamefont {Rico},\ and\ \citenamefont {Kitaev}}]{area_law_7}%
  \BibitemOpen
  \bibfield  {author} {\bibinfo {author} {\bibfnamefont {G.}~\bibnamefont {Vidal}}, \bibinfo {author} {\bibfnamefont {J.~I.}\ \bibnamefont {Latorre}}, \bibinfo {author} {\bibfnamefont {E.}~\bibnamefont {Rico}},\ and\ \bibinfo {author} {\bibfnamefont {A.}~\bibnamefont {Kitaev}},\ }\bibfield  {title} {\bibinfo {title} {Entanglement in quantum critical phenomena},\ }\href {https://doi.org/10.1103/PhysRevLett.90.227902} {\bibfield  {journal} {\bibinfo  {journal} {Phys. Rev. Lett.}\ }\textbf {\bibinfo {volume} {90}},\ \bibinfo {pages} {227902} (\bibinfo {year} {2003})}\BibitemShut {NoStop}%
\end{thebibliography}%

\end{document}